\documentclass[nofootinbib,amsmath,notitlepage,amssymb,preprintnumbers]{revtex4-1}

\usepackage{graphicx,epsfig,psfrag,bm,amssymb}
\usepackage{amssymb,esvect,amsmath,graphicx,latexsym,amsthm,slashed,eso-pic}
\usepackage{dcolumn}
\usepackage{bm}
\usepackage{multirow}
\usepackage{hyperref}

\usepackage{color}
\usepackage{mathrsfs,amsfonts,hepunits, color}

\newcommand{\be}{\begin{eqnarray}}
\newcommand{\ee}{\end{eqnarray}}

\newcommand\FNAL{Fermi National Accelerator Laboratory, Batavia, IL, 60510, United States}
\newcommand{\Eq}[1]{Eq.~(\ref{#1})}

\begin{document}
\hspace{13cm}{FERMILAB-PUB-19-271-A} \\

\title{New Limits on Charged Dark Matter from Large-Scale Coherent Magnetic Fields}
\date{\today}
\author{Albert Stebbins}
\author{Gordan Krnjaic}

\affiliation{\FNAL}

\begin{abstract}
We study the interaction of an electrically charged component of the dark matter with a magnetized galactic interstellar medium (ISM) of (rotating) spiral galaxies.  For the observed ordered component of the field, $B\sim \mu$G, we find that the accumulated Lorentz interactions between the charged particles and the ISM will extract an order unity fraction of the disk angular momentum over the few Gyr Galactic lifetime unless $q/e \lesssim 10^{-13\pm1}\,m\,c^2/\GeV$ if all the dark matter is charged. The bound is weakened by factor $f_{\rm qdm}^{-1/2}$ if only a mass fraction $f_{\rm qdm}\gtrsim0.13$ of the dark matter is charged.  Here $q$ and $m$ are the dark matter particle mass and charge.  If $f_{\rm qdm}\approx1$ this bound excludes charged dark matter produced via the {\it freeze-in} mechanism for $m \lesssim$ TeV/$c^2$.  This bound on $q/m$, obtained from Milky Way parameters, is rough and not based on any precise empirical test.  However this bound is extremely strong and should motivate further work to better model the interaction of charged dark matter with ordered and disordered magnetic fields in galaxies and clusters of galaxies;  to develop precise tests for the presence of  charged dark matter based on better estimates of angular momentum exchange; and also to better understand how charged dark matter might modify the growth of magnetic fields, and the formation and interaction histories of galaxies, galaxy groups, and clusters.
\end{abstract}

\maketitle




\section{Introduction}

\label{intro}
There is compelling evidence for the existence of dark matter (DM) from various independent observations including galactic rotation curves \cite{Sofue:2000jx}, gravitational lensing \cite{Massey:2010hh}, the cosmic microwave background \cite{Aghanim:2018eyx}, and the matter power spectrum \cite{Primack:2015kpa}. However, this body of knowledge is based purely on DM's large scale ($\gtrsim$ kpc) gravitational interactions; its microscopic properties are currently unknown and consistent with a vast landscape of theoretical possibilities -- see \cite{Bertone:2016nfn} for a historical survey. Thus, understanding DM's possible non gravitational interactions on smaller scales is a key priority in fundamental physics.

One minimal, highly predictive possibility is that DM is a fundamental particle with an electric charge $q$ which might be quite small, $q\ll e$ (``millicharged"), or fairly large $q\sim e$, depending on the DM mass.  While a unit electric charge for a dominant DM species is excluded if its mass is not very large, DM particles with unit charge and very high mass or intermediate mass particles with feeble charges remain viable -- see \cite{Davidson:2000hf} for a review. Here we remain agnostic about the preferred size of $q$ and refer to this hypothetical scenario as qDM, as opposed to millicharged or  ``WIMPzilla" DM \cite{Kolb:1998ki}.

In this paper, we consider the effect of qDM on the interstellar medium (ISM) disk of spiral galaxies via their interaction with the embedded ordered magnetic fields to obtain a strong new limit on the DM charge-to-mass ratio.  Charged  particles passing through the disk of a spiral galaxy are deflected by the magnetic fields and thereby exchange momentum between the rapidly rotating ISM and the non-rotating (or more slowly rotating) qDM halo -- for a review of ISM physics see \cite{Ferriere:2001rg}. This angular momentum loss will be shared with all of the ISM gas through acoustic waves, Alfven waves, etc., causing the entire ISM disk to spiral inward as it radiatively emits the excess gravitational energy acquired during its contraction.  By contrast, the stellar disk is largely unaffected by the presence of qDM, so for sufficiently large $q/m$, the ISM ultimately becomes shrunken and embedded in a much larger stellar disk. Since such ISM contraction is not observed in typical spiral galaxies, including our own, this argument places stringent limits on $q/m$. Intriguingly, we find that for qDM masses below $\sim$ TeV$/c^2$, our bound constrains much of the theoretically appealing parameter space predicted by qDM ``freeze-in" production through electromagnetic interactions with Standard Model particles in the early universe \cite{Dodelson:1993je,Hall:2009bx,Chu:2011be,Dvorkin:2019zdi}.    

The origin of the bound and the nature of the interaction are easiest to understand in the so-called {\it diffusive} regime in which $R_{\rm L} \ll \ell_B$, where $\ell_B \sim$ kpc is  the  coherence length of the galactic $B$ field and
\be
 R_{L}\equiv \frac{ m\,c\,v}{q\,B} \simeq 1\,{\rm kpc}  \left( \frac{10^{-12}}{q/e} \right)   \left( \frac{m\,c^2}{{ \rm GeV}} \right) \left( \frac{v/c}{10^{-3}} \right) \left( \frac{\mu \rm G}{B} \right) ,
\ee
is the Larmor radius in CGS units, where $m$ is the mass of the qDM population and $q$ is its charge.  In this regime, qDM particles passing through the ISM quickly lose ``memory" of their initial (vector) velocity and eventually exit the disk with a (vector) velocity, which on average is at rest with respect to the ISM, even though a typical qDM particle's speed does not  change appreciably in a single crossing.  Thus, the qDM halo absorbs an order-unity fraction of the ISM angular momentum once the total DM mass impacting the disk is comparable to the mass contained within the ISM disk. For a representative ISM surface mass density $\Sigma_{\rm ism} \sim 10\, M_\odot/{\rm pc}^2$, qDM mass density $\rho_{\rm qdm}\sim  10^{-2} \, M_{\odot}/{\rm pc}^3$, and a qDM-ISM relative velocity $\sim 250\,{\rm km/s}$,  this occurs in approximately $\sim 10^6$ yr; an extremely short timescale for a galaxy!  Since such an {\it ISM spin-down} phenomenon has not been observed to occur and must be avoided.  Avoiding the diffusive regime requires $\ell_B \lesssim R_{\rm L}$,
 which imposes the approximate bound
\be
\label{diffusiveBound}
\frac{q}{e}\lesssim\frac{v}{c}\,\frac{1}{B\ell_{B}}\sim10^{-12}\, \left( \frac{m\,c^2}{\rm GeV} \right)\,\left( \frac{\mu \rm G}{B} \right)\, 
\left( \frac{\rm kpc}{\ell_B} \right)\,\left(\frac{v}{250\,{\rm km/s}} \right) .
\ee 
Since the $\sim 10^6{\rm yr}$ timescale for spinning down the ISM for this $q/m$ value is so much shorter than the characteristic $>10^9 \, {\rm yr}$ age of a galaxy, one expects more stringent constraints could be found by considering longer-duration effects of particles with even smaller $q/m$ values; we estimate these more stringent bounds below in Sec. \ref{summary}.   Note that the large quantity which drives this stringent bound is the characteristic ``voltage" of the ordered ISM field,  $ \Phi \sim B\ell_{B}\sim10^{18} \, {\rm Volt}$ which must be much less than the ratio $m\,c^2/q$ which also has dimensions of voltage.   Even though magnetic fields do not accelerate particles to higher energies, the momentum they transfer, nonetheless, imposes a very powerful constraint on charged dark matter.  As we will see, for much of the mass range considered in the remainder of this paper, the bound we derive is much stronger than other published bounds on qDM and illustrates the power of considering DM interactions with large scale ordered magnetic fields.   Note that this ISM spin-down constraint does not hold if the total mass of qDM in the halo is smaller than the mass in the ISM.  In this case, instead of the ISM being {\it spun-down}, the qDM would be {\it spun-up}.  For this reason our Solar radius spin-down bound no longer apply if the qDM mass fraction is  $\lesssim 10$\% of the total halo population.

This paper is organized as follows: Section \ref{summary}, presents a summary of the argument presented here; Section \ref{interactions} derives the qDM-ISM momentum transfer rate; Section \ref{galaxy} computes the main constraint on qDM using  local Galactic parameters; Section \ref{best} discusses other places qDM interactions with large scale magnetic fields might  manifest; Section \ref{models} discusses the theoretical implications of various charged DM models to which our bounds apply; and Section \ref{conclusion} offers some concluding remarks. Finally,  Appendix \ref{appendix-small-charges}  discusses how a millicharge may arise in the presence of a hidden photon and Appendix \ref{appendix-freezein} summarizes
millicharge DM freeze-in production




\section{Estimating Spin-Down Limits on Charged Dark Matter}
\label{summary}

Here we outline the derivation a much tighter bound on $q/m$ relative to Eq.~(\ref{diffusiveBound}) obtained by a careful quantitative analysis of the ballistic limit, $R_{L}\gg\ell_{B}$, which is opposite extreme of the diffusive limit.   Allowing the spin-down to proceed over the full $\sim$ few Gyr age of the Galaxy.  In addition to qDM properties, $q$, $m$ and $f_{\rm qdm}\equiv \rho_{\rm qdm}/\rho_{\rm dm}$, which is the mass fraction of dark matter which is charged, we need quantities that describe local properties of the disk and halo: $\rho_{\rm dm}$ ($10^{-2}\,M_\odot/{\rm pc^3}$) is the local dark matter density \cite{Bovy:2012tw}, $\sigma$ ($10^{-3} c$) is the dark matter 1-D velocity dispersion, $\Sigma_{\rm ism}$ ($10\,M_\odot/{\rm pc^2}$) is the ISM surface mass density \cite{Naab:2005km}, $B$ ($\mu{\rm G}$) is the relevant component of coherent disk magnetic field \cite{sobey}, $\Delta z$ ($1\,{\rm kpc}$) is the magnetic field scale height \cite{sobey}, $V$ the circular velocity of the ISM ($250\,{\rm km/s})$, and $T_{\rm ism}$ ($5\,{\rm Gyr}$) the length of time over which the ISM (gas and magnetic field strength) has not changed significantly \cite{Pakmor2017}.   The parenthetical values are appropriate for the Galaxy near the Sun and yield the numerical limits in the following equations. 

\begin{figure*}
\hspace{-0.2in}
\includegraphics[width=4 in,angle=0]{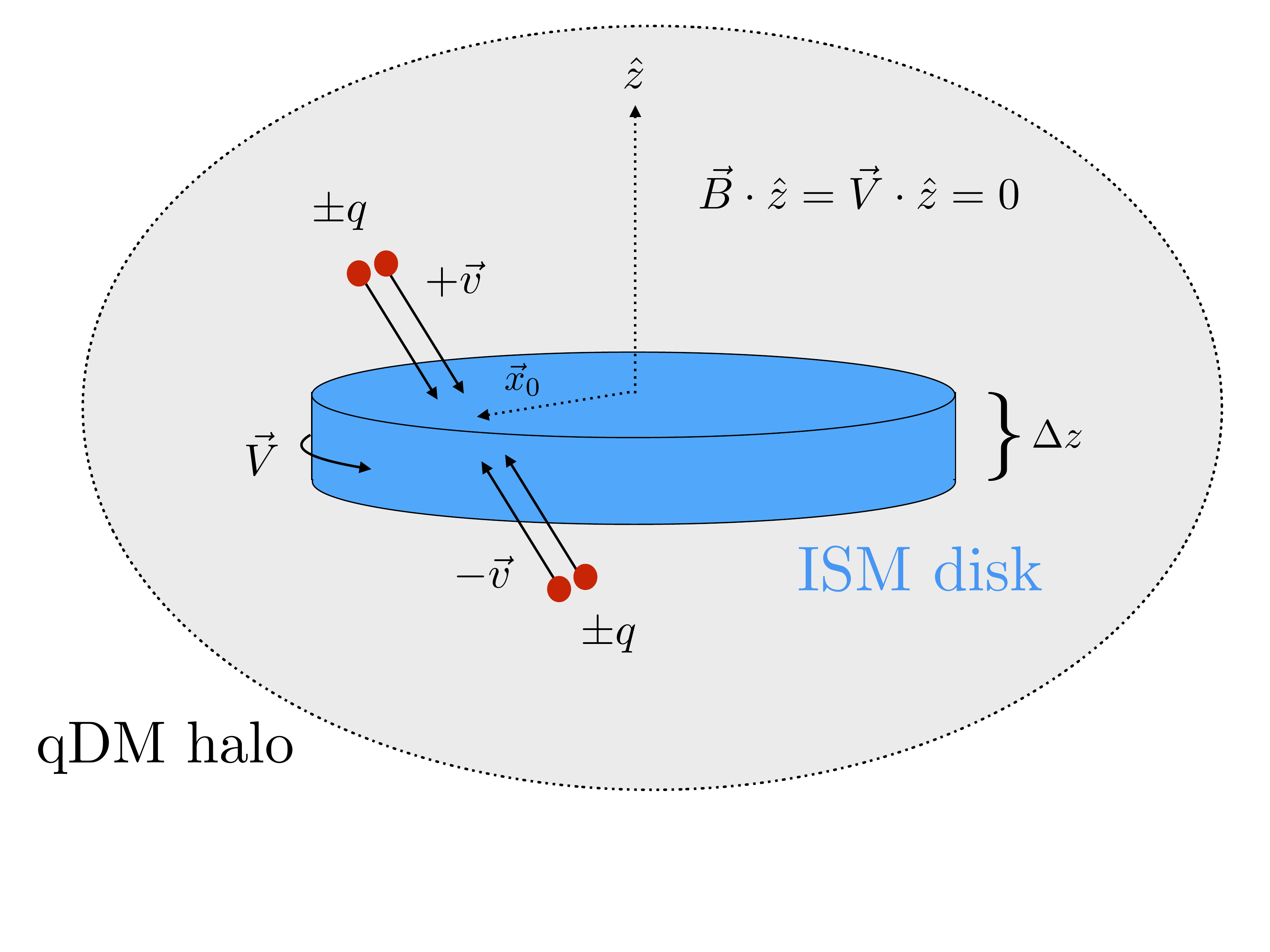}~~~~
\caption{ Schematic cartoon representing the ISM disk model presented in this paper. Here the central ISM disk is modeled as a  uniform, cylindrical gas of thickness $\Delta z \simeq $ kpc, with velocity $\vec V$ and a coherent magnetic field $B \sim \mu$G both in the $xy$ plane. The  disk is immersed in a DM halo, which contains charged dark matter (qDM) particles  that exchange momentum with the ISM disk on galactic $\sim$ Gyr timescales.  We estimate the impact of this transfer by considering the average effect of qDM quadruples (depicted as red dots) that pass through the slab at some representative local point $\vec x_0 \sim R_\odot \sim$ few kpc in the plane of the disk. The combined momentum transfer of opposite-charge pairs $\pm q$ with similar velocities $\sim \vec v$ cancels to leading order in the Born approximation, but yields a nonzero effect at second order. In our treatment, this surviving second order contribution is then averaged with that of a corresponding charge-pair with a mirror reflected velocity $-\vec v$ to obtain the net momentum transfer rate at position $\vec x_0$. This rate is then convolved with a local Maxwellian velocity distribution for the to obtain limits on the $q/m$ ratio and fractional abundance of the qDM population.
}
\label{cartoon}
\end{figure*}

Below in Sec. \ref{interactions} we find that, for these reference Galactic values, the mean momentum transfer of ISM rotational momentum to the qDM per ISM crossing is approximately
\be
\frac{ \Delta p}{mV} \sim \left( \frac{\Delta z}{R_{L}} \right)^2 
\sim \left( \frac{q\,B\Delta z }{m\, c \, \sigma} \right)^2  \lesssim 10^{-4}\, \left( \frac{q}{10^{-14}\, e} \right)^2  \left( \frac{\rm GeV}{m \, c^2} \right)^2,
\ee
where we have averaged over {\it two pairs} of opposite sign charges passing through the local Galaxy as depicted schematically in Fig. \ref{cartoon}.
Multiplying by the qDM particle flux $\sim\rho_{\rm qdm}\,\sigma/m$ gives the total rate of momentum transfer to the qDM from which one can compute the ISM deceleration, 
$\Sigma_{\rm ism}\,\dot{V}\sim-\rho_{\rm qdm}\,\sigma\,\Delta p/m$.  The mechanism by which the ISM is decelerated is a $\vec{J}_{\rm qdm}\times\vec{B}$ force where $J_{\rm qdm}$ is the electric current which results from the differing velocity kicks given to positive and negative charges by $\vec{B}$.  We should be clear that $\dot{V}$ is only the net ISM deceleration by $\vec{J}_{\rm qdm}\times\vec{B}$. Through its interactions with the surrounding gas, the ISM material should be heated and shed excess kinetic energy via radiative cooling, thereby equilibrating to roughly circular orbits at smaller radii. The final circular speed of a patch of gas may be smaller or larger than the initial speed depending on the rotation curve; for a halo with a flat rotation curve the speed would not change at all.

Defining the spin-down timescale by 
\be
\tau_{\rm sd}\equiv-\frac{V}{\dot{V}}
\sim\frac{\Sigma_{\rm ism} }{f_{\rm qdm}\,\rho_{\rm qdm}\,\sigma}\,
\left(\frac{m\,c\,\sigma}{q B\Delta z}  \right)^2\ ,
\ee
we set the criteria for no significant spin-down to be $\tau_{\rm sd}<T_{\rm ism}$.  Using the above formulae but with the numerical factors from our detailed derivation and parameters from the local Galaxy culminates in Eqs. ~(\ref{iterative-spin-down-bound})  and ~(\ref{local-spin-down-bound}), which jointly imply the bound
\be
\label{limitPreview}
\sqrt{f_{\rm qdm}}\,\frac{q}{e}  \lesssim 
\sqrt{\frac{2\sqrt{2\pi}}{1-0.2\,{\rm ln}f_{\rm qdm}}}\,
\frac{m\,c^2}{B\Delta z}\,
\sqrt{\frac{\Sigma_{\rm ism}}{\rho_{\rm dm} c \, T_{\rm ism}} \frac{\sigma}{c}} 
~\sim~  10^{-14}\left( \frac{m c^2}{\rm GeV} \right)
\quad {\rm when}\quad f_{\rm qdm}\gtrsim 0.1
 ~,~~
\ee
which is nearly two orders of magnitudes stronger than Eq.~(\ref{diffusiveBound}) for $f_{\rm qdm}=1$; this result is summarized  graphically in Figs. \ref{qDMlimits} and \ref{millichargeBounds}.  Note that this bound is not on $q/m$ but rather on $\sqrt{f_{\rm qdm}}\,q/m$; and only applies when $f_{\rm qdm}\gtrsim M_{\rm ism}/M_{\rm qdm}$ where $M_{\rm ism}\sim\Sigma_{\rm ism} R^2$ is the inertial mass of the ISM, $M_{\rm qdm}\sim f_{\rm qdm}\,\rho_{\rm dm}\,R^3$ is the inertial mass of the qDM passing through the disk and $R$ is the galacto-centric  radius.  For smaller $f_{\rm qdm}$ the qDM subcomponent will be {\it spun-up}; its velocity distribution will be significantly modified, but the ISM will remain largely unaffected.

The upper bound on $q/m$ in Eq.~(\ref{limitPreview}) differs from that of Eq.~(\ref{diffusiveBound}) by two additional factors which may be written
$\sqrt{M_{\rm ism}/M_{\rm qdm}}$ $\times$ $\sqrt{1/N_{\rm cross}}$
where $N_{\rm cross}\sim \sigma\,T_{\rm ism}/R$ is the number of times a typical qDM particle crosses the disk in time $T_{\rm ism}$.  It is clear that both of these factors will strengthen the bound of Eq.~(\ref{limitPreview}).  The reason the factors enter with a square root has to do with the fact that the momentum transfer is only nonzero to 2nd order in a perturbative expansion, $\Delta p\propto(\Delta z/R_{L})^2$ (the 2nd Born approximation).

As elaborated in Sec. \ref{galaxy}, we set limits using Solar neighborhood parameters in our numerics simply because we have a better observational indications of our local environment and not because the spin-down effect is maximized here.  Even so, the observational uncertainties in these quantities, especially $B$, lead to more than an order of magnitude uncertainty in an observational bound on $\sqrt{f_{\rm qdm}}\,q/m$. Parametric uncertainties are liable to be larger in other parts of our Galaxy or in other galaxies where one also might set bounds on $q/m$.  In spite of these larger uncertainties, observational bounds from other parts of the universe might be stronger than the local Solar bound we use here.  In particular, regions more dark matter dominated than the Solar neighborhood could potentially set stronger limits on $f_{\rm qdm}$.  In addition to parametric uncertainties, one must also consider uncertainties associated with the implementation of an observational test to determine whether ISM spin-down has occurred in a given galactic system; however, in the present work, we do not specify such a test. For this reason we allow a full two orders-of-magnitude in additional uncertainty in Figs.~\ref{qDMlimits} and \ref{millichargeBounds}, which summarize our formal results for the Solar neighborhood.  The large uncertainties in the bound could be reduced by both more accurate determination of the local magnetic field pattern and by particle+magnetohydrodynamical simulations to clarify which observational measurements best restrict the amount of spin-down.

\section{Interaction of ISM with Charged Dark Matter in the Halo}
\label{interactions}

\subsubsection{Assumptions}

\begin{figure}[t!]
\hspace{-0.2in}
\includegraphics[width=7in,angle=0]{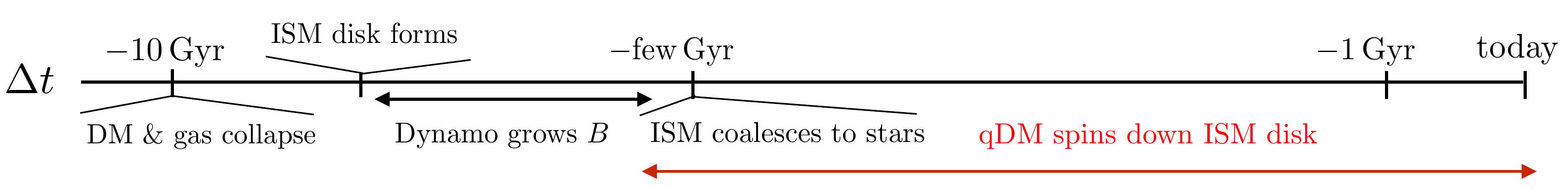}~~~~
\caption{Timeline of events relevant for qDM induced spin-down. 
}
\label{timeline}
\end{figure}

We now begin a more formal derivation of the rate of ISM spin-down.  Our particle model assumption is 
\begin{enumerate}
\item{} There is only one qDM particle species of mass $m$ with equal numbers of charges $\pm q$ which makes up a mass fraction $f_{\rm qdm} \le 1$ of all the dark matter and only interacts with Standard Model particles through gravitational and electromagnetic forces.
\end{enumerate}
and since we are working in the ballistic limit we are assuming
\begin{enumerate}
\setcounter{enumi}{1}
\item{} the typical Larmor radius of qDM passing through the disk is much greater than the disk thickness 
($R_{L}\sim m\,c\,\sigma/q\,B\gg\Delta z$), so qDM particle trajectories are only slightly perturbed by $\vec{B}$ (the Born approximation). \end{enumerate}
We expand to 2nd order in $\Delta z/R_{L}$ (2nd Born approximation) to obtain the leading-order contribution to ISM spin-down and obtain our spin-down bound.  To model spin-down precisely in the 2nd Born approximation we make a number of additional assumptions.  Some of these are likely accurate:
\begin{enumerate}
\setcounter{enumi}{2}
\item{} non-relativistic dynamics,
\item{} the MHD approximation,
\item{}  the magnetic fields are quasi-static, i.e. magnetic plasma waves move slowly compared to typical qDM velocities (see \cite{Klessen2015} which reviews ISM plasma dynamics);
\end{enumerate}
and some are more dubious and might be made differently
\begin{enumerate}
\setcounter{enumi}{5}
\item{} the disk ISM and magnetic fields are formed and stable before the effects of qDM interactions become important.
\item{}  the magnetic fields are concentrated near the disk ($\Delta z\ll R$).
\item{} the gravitational deflection of qDM orbits is negligible during a passage through the ISM.
\end{enumerate}
To derive a numerical bound on $q/m$ we also assume
\begin{enumerate}
\setcounter{enumi}{8}
\item{} the dark matter halo is rotating much more slowly than the ISM,
\item{} the ISM  and embedded ordered magnetic fields orbit the galaxy center in nearly circular trajectories,
\item{} current local values of ISM and DM properties are representative of their values over cosmic time,
\item{} the local magnetic field pattern can be adequately modeled with planar symmetry,
\item{} various assumptions on  how to interpret data probing the local magnetic field 
\end{enumerate}
The timeline of assumed events is given in Fig~\ref{timeline}.  It is interesting to explore whether qDM interactions with magnetic fields would actually modify galaxy formation and/or the growth of magnetic fields and violate this timeline,  but such an inquiry is beyond the scope of this paper.

\subsubsection{Lorentz Force in Quasi-static MHD Approximation}

The fundamental interaction we consider is the usual electromagnetic Lorentz force 
\be
\label{LorentzForce}
\vec{F}=\pm q\,\left( \vec{E}  + \frac{1}{c}\vec{v}\times\vec B\right)\ ,
\ee
where $q$ is the charge of a qDM particle passing through the ISM disk with velocity $\vec v$.   Assuming the magnetic field is embedded in a highly conducting medium (the ISM) moving with velocity $\vec{V}$, the electrons and ions will create currents to short out any $\vec{E}\cdot\vec{B}\ne0$ so to a good approximation we may use the ideal Ohm's law of the MHD approximation \cite{krall1973principles}
\be
\label{MHD approximation}
 \vec{E}\approx-\frac{1}{c}\vec{V}\times\vec{B},
 \ee
so the Lorentz force law may be rewritten
\be
\label{LorentzForceMHD}
\vec{F}\approx\pm\frac{q}{c}(\vec{v}-\vec{V})\times\vec{B}\ .
\ee
This form of the equations of motion for qDM particles no longer exhibits Galilean invariance since the ISM frame breaks this symmetry.  This is how qDM halo particles ``know" to move toward co-rotation with the ISM.

In order to spin-down the ISM, the qDM must extract rotational energy from the ISM in the galactic rest frame.  One cannot (naively) argue that (purely) magnetic fields cannot change the energy of the qDM particles and therefore cannot lead to an ISM to qDM energy exchange.  Energy, $\vec{E}$, and $\vec{B}$ are each frame-dependent quantities; in one frame there might be energy exchange and another not.  Indeed, in the MHD approximation ($\vec{E}\cdot\vec{B}=0$), whenever $|\vec{B}|>|\vec{E}|$  there always exists a local frame in which $\vec{E}=\vec{0}$ and charged-particle kinetic-energy is conserved.  Our MHD approximation sets the $\vec{E}=\vec{0}$ frame to be the ISM frame and since this is different from the galactic rest frame rotational energy may be extracted in the galactic frame; Eq.~(\ref{LorentzForceMHD}) does hide the fact that $\vec{E}\ne\vec{0}$ in other frames.  The rate of energy transfer to a qDM particle is 
$\dot{\epsilon}_\pm=\vec{v}\cdot\vec{F}\approx\mp\frac{q}{c}\,\vec{v}\cdot(\vec{V}\times\vec{B})$
which is usually non-zero. 

\subsubsection{Maxwell-Vlasov Equation}
 
While in what follows we work with the microscopic description of single particle trajectories, this treatment is equivalent to a more macroscopic (non-relativistic) Maxwell-Vlasov equation which, with our MHD approximation, is
\be
\frac{\partial}{\partial t}\,f_\pm+\vec{v}\cdot\frac{\partial}{\partial\vec{x}}\,f_\pm
=\pm\frac{q}{m\,c}\left[(\vec{v}-\vec{V})\times\vec{B}\right]\frac{\partial}{\partial\vec{v}}\,f_\pm\ ,
\ee
where the qDM particles are described by a statistical distribution function $f_\pm(\vec{x},\vec{v})$ for positively ($+$) and negatively ($-$) charged particles, which is normalized such that the number density of each species is
$n_\pm\equiv\int d^3\vec{v}\,f_\pm$.   We assume that initially $f_+=f_-=f_0$, but $f$ will evolve away from this as the qDM particles travel through the magnetic field.  This can result in a qDM electric charge and current density, 
$\varrho_{\rm qdm}\equiv q\,(n_+-n_-)$ and $\vec{J}_{\rm qdm}\equiv q\,\int d^3\vec{v}\,\vec{v}\,(f_+-f_-)$ respectively, both of which are initially zero.  In this macroscopic description the force density on the qDM is 
$\varrho_{\rm qdm}\,\vec{E}+\vec{J}_{\rm qdm}\times\vec{B}/c$ with an equal and opposite force applied to the conducting ISM in which the fields are embedded.  Since our calculation is a perturbative expansion in $\Delta z/R_{L}\propto |\vec{B}|$, to lowest order we should use the unperturbed ISM $\vec{B}$ and, since the unperturbed $\vec{E}$ is zero in the ISM frame, we can think of the force as a $\vec{J}_{\rm qdm}\times\vec{B}$ force.

\subsection{Momentum Transfer}

Returning now to a microscopic point-particle treatment:  from Eq.~(\ref{LorentzForceMHD}), charged qDM particles of mass $m$ and charge $\pm q$ will follow trajectories
\be
\label{xddot}
\ddot{\vec x}_\pm(t) = \pm \frac{q}{m c} \left( \dot{\vec{x}}_\pm(t) - \vec{V}   \right)   \times \vec B[\vec x_\pm(t)]\ .
\ee
Assuming the particle eventually exits the disk, the momentum transfer it acquires is formally given by
\be
\label{delta-p-master0}
\Delta \vec{p}_1^{\,\pm}
= m \int_{-\infty}^\infty dt\,\ddot{\vec{x}}_\pm(t)
= \pm\frac{q}{c}\int_{-\infty}^{+\infty} dt\,\left(\dot{\vec{x}}_\pm(t)-\vec{V}\right)\times\vec{B}[\vec{x}_\pm(t)]~,
\ee
where the 1 subscript indicates that this is a single-particle quantity.  This formal result is not useful unless one knows $\vec{x}_\pm(t)$, which we obtain below from a perturbative expansion in the field strength $B=|\vec{B}|$.

\subsubsection{Born Approximations}

We wish to compute the spin-down effect in the so-called {\it ballistic} limit in which  particle trajectories are only slightly deflected from their $q=0$ paths by $\vec{B}$ as they  traverse the ISM assuming that the gravitational deflections are negligible over this short $\sim$ kpc distance.  For slight deflections, the time it takes to traverse the ISM of thickness $\Delta z$ is $\sim\Delta z/v$ where $v\gtrsim V$ is the particle's speed.  For a single particle, the fractional change in momentum (or velocity) is
$\Delta p/p\lesssim (q\,B\,\Delta z/c)/(m\,v)=\Delta z/R_{L}$
where $R_{L}= m\,v\,c/(q\,B)$ is the Larmor radius.  So a sufficient condition for small deflections is $\Delta z\ll R_{L}$.   This is the opposite of the diffusive limit ($\ell_{\rm B} \gg R_{L}$) since $\Delta z\gtrsim\ell_{\rm B}$.

We expand in the small parameter $\Delta z/R_{L}$ and use $X^{(p)}$ to denote any quantity $X$ accurate to $(\Delta z/R_{L})^p$ and use {\it order} to refer to the value of $p$.  The Born approximation usually means one uses the undeflected trajectory but we use Born for all orders: $p=1$ is the 1st Born Approximation and $p=2$ the 2nd Born Approximation.  Since Eq.~\ref{delta-p-master} has a $\vec{B}$ in it one can use the trajectory $\vec{x}^{(p-1)}_\pm$ (order $p-1$) to compute momentum transfer $\Delta\vec{p}^{\,(p)}$ (order $p$).

 \subsubsection{Momentum Transfer in the 1st Born Approximation}

To lowest, zeroth, order the trajectories are unperturbed by $\vec{B}$ and given by
\be
\label{zerotheom}
\vec{x}^{(0)}_\pm(t) = \vec{x}_0 + \vec{v}\,(t-t_0) ~~~,~~~ \dot{\vec{x}}^{(0)}_\pm = \vec{v}~,
\ee
where $\vec{v}$ is the initial (unperturbed) velocity while, $t_0$ and $\vec{x}_0$ give the time and position where the unperturbed trajectory would have crossed the central plane of the ISM ($z=0$)  so $\hat{z}\cdot\vec{x}_0=0$.  For notational simplicity we suppress the $\vec{x}_0$ dependence in most formulae as we are considering only a small region of the disk.

To lowest, 1st, order the momentum transfer to a single particle is obtained by substituting Eq.~(\ref{zerotheom}) into Eq.~(\ref{delta-p-master0}) 
\be
\label{delta-p-1}
\Delta \vec{p}_1^{\,\pm\,(1)}
= \pm\frac{q}{c}\int_{-\infty}^{+\infty} d\tau\,\left(\vec{v}-\vec{V}\right)\times\vec{B}(\vec{x}_0+\vec{v}\,\tau)~,
\ee
where $\tau \equiv t - t_0$.  Note that all quantities are independent of $t_0$ since $\vec B$ is assumed to vary slowly enough to be considered time independent on the timescale of a qDM particle crossing.

It is convenient  to average the momentum transferred to an oppositely charged pairs of qDM particles starting on the same trajectory since to the extent they remain on the same trajectory the Lorentz force on the pair sums to zero and there is no net momentum transfer.  We are not supposing two literal particles start on the same trajectory but rather we are assuming it is equally likely that a positive or negative charge start on any particular trajectory. The mean momentum transferred to the pair of oppositely charged particles is 
\be
\label{delta-p-master}
\Delta \vec{p}_2=\frac{1}{2}\sum_\pm \Delta \vec{p}_1^{\,\pm}~,
\ee
where the factor of $1/2$ comes from averaging the $\pm$ contributions and the subscript 2 indicates this quantity represents a two-particle average. So $\Delta \vec{p}_2$ to lowest order is
\be
\Delta\vec{p}_2^{\,(1)}= \frac{1}{2}\sum_\pm\Delta\vec{p}_1^{\,\pm\,(1)}= 0,
\ee
and one must go to higher order to find the leading order contribution to spin-down.

\subsubsection{Second Born Approximation}

Since Pharaoh took the first Born, we rely on the second Born to avenge this injustice, which is the approximation
\be
\vec{x}^{\ (1)}_\pm(\tau) = \vec x_0 + \vec v \tau \pm \delta \vec{x}(\tau)  ~~~,~~~
\dot{\vec{x}}^{\ (1)}_\pm(\tau) = \vec v \pm \delta \dot{\vec{x}}(\tau),
\ee
 so using Eq.~(\ref{delta-p-master0}), the leading order momentum transfer is now 
\be
\label{deltap2}
\Delta\vec{p}_2^{\ (2)} ({\vec v})&=& \frac{q}{2c} \sum_\pm \pm \int_{-\infty}^\infty d\tau (\dot{\vec{x}}^{\ (1)}_\pm - \vec V ) \times \vec B(\vec x_0 + \vec v \tau)
= \frac{q}{c} \int_{-\infty}^\infty d\tau \,\delta \dot{\vec{x}}(\tau)  \times \vec B(\vec x_0 + \vec v \tau) ,
 \ee
 where, in the last equality all terms independent of $\delta \dot{\vec{x}}$ cancel when we include the contributions from 
 $\pm$ charges.   Since the surviving piece is proportional to $q\,\delta \dot{\vec{x}}$, both charge species
 receive the same momentum transfer if their initial, unperturbed trajectories are similar, as stipulated in Eq.~(\ref{zerotheom}).
The first order velocity perturbation is obtained by integrating Eq.~(\ref{xddot}), so we have 
\be
\label{second-born-velocity}
\delta\dot{ \vec{x}}(\tau) \equiv \frac{q}{m\,c}(\vec v - \vec V) \times \int_{-\infty}^\tau d\tau^\prime \vec B(\vec{x}_0 + \vec v \tau^\prime)~,
\ee
 which can be substituted into Eq.~(\ref{deltap2}) to yield
 \be
\Delta\vec{p}_2^{\ (2)} ({\vec v})&=& \frac{q^2}{m\,c^2} \int_{-\infty}^\infty d\tau \, \int_{-\infty}^\tau d\tau^\prime\,
\left[   (\vec v - \vec V) \times\vec B(\vec{x}_0 + \vec v \tau^\prime)\right]\times \vec B(\vec x_0 + \vec v \tau) 
  \\
&=& \frac{q^2}{m\,c^2} \int_{-\infty}^\infty d\tau \, \int_{-\infty}^\tau d\tau^\prime\,
\left[
-\vec B(\vec{x}_0 + \vec v \tau)\cdot\vec B(\vec x_0 + \vec v \tau^\prime)\,(\vec v-\vec V)
+(\vec v - \vec V)\cdot \vec  B(\vec{x}_0 + \vec v \tau)\, \vec B(\vec x_0 + \vec v \tau^\prime)
\right] ,  \label{deltap2sub} 
\ee
where we have used the identity
$(\vec a \times \vec b \, )\times \vec c = -(\vec b \cdot \vec c \,) \vec a+ (\vec a\cdot \vec c \, ) \vec b$.  
Defining the tensors 
$ V^{ijk}\equiv V^i \delta^{jk}-V^j\delta^{ik}$ and $ v^{ijk}\equiv v^i \delta^{jk}-v^j\delta^{ik}$,
  we can rewrite Eq.~(\ref{deltap2sub}) in component form as
\be
\label{deltap2final}
[\Delta\vec{p}_2^{\ (2)}({\vec v})]^i =\left(V^{ijk}-v^{ijk}\right)
\frac{q^2}{m\,c^2\,v^2}\,\Phi^2_{jk}(\hat{v}) ~~~, ~~~
\Phi^2_{jk}(\hat{v})\equiv\int_{-\infty}^\infty du \, \int_{-\infty}^{\> u} du^\prime\,
B_j(\vec{x}_0 + \hat{v}u)\,B_k(\vec x_0 + \hat{v}u^\prime)\ ,
\ee
where $v\equiv |{\vec v}|, \hat{v}\equiv\vec{v}/v$, and integration variables have been changed from time to distance $\tau\to u\equiv v\tau$, $\tau^\prime \to u^\prime\equiv v\tau^\prime$. Note that  $\Phi^2_{jk}$ has dimensions of voltage squared and is generically very large for typical ISM parameters, as discussed previously. Eq.~(\ref{deltap2final}) gives the leading order momentum transfer from which we derive our qDM limit.  In what follows, we merely try to simplify this expression.  

\subsubsection{Four Particle Average of Momentum Transfer}

To simplify our main result further, we average the momentum transfer over {\it two pairs} of particles (4 particles total) moving in opposite directions, $\pm\vec{v}$ but intersecting the central plane at the same location $\vec{x}_0$ as depicted schematically in Fig.~\ref{cartoon}.  We need not assume the flux of particles from opposite directions is exactly the same although in the galaxy frame we expect the qDM distribution function to be nearly $\pm\vec{v}$ symmetric apart from rotation of the dark matter halo and any infalling DM substructure.  The 4 particle momentum transfers we need are
\be
\Delta\vec{p}_{4\pm}^{\ (2)}({\vec v})\equiv
\frac{1}{2} \left[ \Delta\vec{p}_2^{\ (2)}({\vec v})\pm\Delta\vec{p}_2^{\ (2)}({-\vec v})\right]\ ,
\ee
where subscript 4 indicates a 4 particle average and in this subsection the $+$ and $-$ do  {\it not} refer to the sign of the charge but rather whether to add or subtract the momentum transfer of the oppositely moving pairs.  The $(+)$ reflection symmetric term gives the average of the two pairs and an additional $(-)$ reflection antisymmetric term allows one to weight the $\pm\vec{v}$ pairs differently which one will need if there is more qDM moving in one direction than in the opposite direction.


Since $V^{ijk}$ and $v^{ijk}$ are respectively symmetric and anti-symmetric under $\vec v \to -\vec v$ we will need both the symmetric and anti-symmetric parts of $\Phi^2_{jk}(\hat{v})$ for symmetric and anti-symmetric distributions, which are given by
\be
\Phi^{2\pm}_{jk}(\hat{v})=\frac{1}{2}\left[\Phi^2_{jk}(+\hat{v})\pm\Phi^2_{jk}(-\hat{v})\right]~.
\ee
After changing the variables in Eq.~(\ref{deltap2final}) for the $\Phi^2_{jk}(-\hat{v})$ integral, $u \to -u$ and $u^\prime \to -u^\prime$, the integration range for the sum of the two terms expands to cover the entire $u$-$u^\prime$ plane and one finds
\be
\Phi^{2+}_{jk}(\hat{v})&=&{1\over2}\int_{-\infty}^\infty du \, \int_{-\infty}^\infty du^\prime\,
B_j(\vec{x}_0 + \hat{v}\,u)\,B_k(\vec x_0 + \hat{v}\,u^\prime) \\
\Phi^{2-}_{jk}(\hat{v})&=&{1\over2}\int_{-\infty}^\infty du \, \int_{-\infty}^\infty du^\prime\,
B_j(\vec{x}_0 + \hat{v}\,u)\,B_k(\vec x_0 + \hat{v}\,u^\prime)\,{\rm sgn}(u-u^\prime)\ ,
\ee
which are respectively symmetric and anti-symmetric 3-tensors: 
$\Phi^{2\pm}_{jk}(\hat{v})=\pm\Phi^{2\pm}_{kj}(-\hat{v})$.  The symmetric $\Phi^{2+}_{jk}$ term factorizes into two 3-vectors
\be
\label{defPhi}
\Phi^{2+}_{jk}(\hat{v})={1\over2}\Phi_j(\hat{v})\,\Phi_k(\hat{v})~~,~~
\vec\Phi(\hat{v})\equiv\int_{-\infty}^\infty du\,\vec{B}(\vec{x}_0 + \hat{v}\,u)
\ee
while the 2nd, being an antisymmetric tensor, has a dual 3-vector
$\Phi^i_2(\hat{v})\equiv 2\epsilon^{ijk}\,\Phi^{2-}_{jk}(\hat{v})$ where $\epsilon^{ijk}$ is the completely anti-symmetric tensor and we have 
\be
\Phi^{2-}_{jk}(\hat{v})={1\over4}\,\epsilon_{ijk}\,\Phi^i_2  ~~,~~
\vec\Phi_2(\hat{v})\equiv
\int_{-\infty}^\infty du \, \int_{-\infty}^\infty du^\prime\,{\rm sgn}(u-u^\prime)\,
\vec{B}(\vec{x}_0 + \hat{v}\,u)\times\vec{B}(\vec x_0 + \hat{v}\,u^\prime)\ .
\ee
Thus, in the 2nd Born approximation, the net momentum transfer dependence on the $\vec{B}$ configuration is given by just two vector-valued quantities, $\vec{\Phi}$ and $\vec{\Phi}_2$ -- note that $\vec{\Phi}$ has dimensions of voltage whereas $\vec{\Phi}_2$ has dimensions of voltage {\it squared}. Only at higher order in the Born approximation will more details of $\vec{B}$ be required to compute momentum transfer.  $\vec{\Phi}$ measures mean $\vec{B}$ along the qDM trajectory and thus only probes magnetic fields on the largest scales. Physically, $\vec{\Phi}_2$ is a measure of the {\it twist} in $\vec{B}$ along the photon trajectory since it is zero unless the direction of $\vec{B}$ varies.  Thus $\vec{\Phi}_2$ is sensitive to smaller scale structure of $\vec{B}$ than does $\vec{\Phi}$ although it is only mildly weighted toward smaller scales.  
$\vec{\Phi}$ has dimensions of voltage and $\vec{\Phi}_2$ voltage squared which can be compared to $q/(m\,c^2)$ which also has dimensions of voltage.  Note that the Born approximation employed in this calculation only applies if $|\vec{\Phi}|\,,|\vec{\Phi}_2|\ll q/(m\,c^2)$. 

The momentum transfer from reflection symmetric/antisymmetric pairs of qDM opposite-sign pairs is
\be
\label{dPdecomposition}
\Delta\vec{p}_{4\pm}^{\ (2)}({\vec v})&=&\frac{q^2}{m\,c^2v^2}\,
\left(V^{ijk}\Phi^{2\pm}_{jk}(\hat{v})-v^{ijk}\Phi^{2\mp}_{jk}(\hat{v})\right)
=\Delta\vec{p}_{\rm V\pm}({\vec v})+\Delta\vec{p}_{\rm v\pm}({\vec v}) ,
\ee
where we have defined the quantities
\be
\Delta\vec{p}_{\rm V+}({\vec v})&\equiv&+\frac{m}{2}\,\left(\frac{q}{m\,c\,v}\right)^2\,
\left(|\vec\Phi(\hat{v})|^2\,\vec{V}
-\vec{V}\cdot\vec\Phi(\hat{v})\,\vec\Phi(\hat{v})\right) ~~~,\quad
\Delta\vec{p}_{\rm v+}({\vec v})\equiv-\frac{m}{4}\,\left(\frac{q}{m\,c\,v}\right)^2\,
\vec{v}\times\vec{\Phi}_2(\hat{v}) \nonumber \\
\Delta\vec{p}_{\rm v-}({\vec v})&\equiv&-\frac{m}{2}\,\left(\frac{q}{m\,c\,v}\right)^2\,
\left(|\vec\Phi(\hat{v})|^2\,\vec{v}
-\vec{v}\cdot\vec\Phi(\hat{v})\,\vec\Phi(\hat{v})\right)  \quad~~,\quad
\Delta\vec{p}_{\rm V-}({\vec v})\equiv+\frac{m}{4}\,\left(\frac{q}{m\,c\,v}\right)^2\,
\vec{V}\times\vec{\Phi}_2(\hat{v}) \nonumber \ ,
\ee
and have used the properties $\vec{\Phi}(\hat{v})=+\vec{\Phi}(-\hat{v})$ and $\vec{\Phi}_2(\hat{v})=-\vec{\Phi}_2(-\hat{v})$.

\subsubsection{Momentum Transfer from a Distribution of Particles}

The rate of momentum per unit area transferred between a distribution of qDM passing through a small patch of a galactic disk ISM is given by integrating the momentum transferred per particle multiplied by the rate of qDM particles passing through the central plane per unit area as a function of velocity: $dR/d^2 \vec{x}_{0} d^3\vec v$.  This latter quantity is related to the unperturbed qDM velocity distribution function $f_0(\vec{v})$ and mass density by
\be
\label{rho}
\frac{dR}{d^2 \vec{x}_0 d^3 \vec{v}} = |v_z|\,f_0(\vec v)~~~, ~~~
\rho_{\rm qdm} =  m \int d^3 \vec{v}\,f_0(\vec v)~,
\ee
where $v_z\equiv\hat{z}\cdot\vec{v}$.  Since we allow for distributions which are not reflection symmetric ($\vec{v}\to-\vec{v}$), we can decompose $f_0$ into symmetric and antisymmetric parts
\be
f_0^\pm(\vec{v})\equiv\frac{1}{2}\left[ f_0(+\vec{v})\pm f_0(-\vec{v})\right]~.
\ee
  Since $f_0\ge0$ we know that $f_0^+\ge0$ and $|f_0^-|\le f_0^+$.  Also,  note that  the integrals of $f_0^\pm$ satisfy
  \be
  \int d^3\vec{v}\,f_0^+(\vec{v})=\frac{\rho_{\rm qdm}}{m}~~,~~~\int d^3\vec{v}\,f_0^-(\vec{v})=0,
  \ee
where the latter integrand is odd under all velocity components. In the 2nd Born approximation the momentum transfer rate per unit area is
\be
\label{momentum-transfer-per-area}
\frac{d}{dt}\frac{d\vec{P}}{d^2\vec x_0}
=\int d^3 \vec{v}\,\left( \frac{dR}{d^2 \vec{x}_0 d^3 \vec{v}} \right)\,\Delta\vec{p}_{4\pm}^{\ (1)}({\vec v})
=\sum_\pm \int d^3 \vec{v}\, |v_z|\,f_0^\pm(\vec{v})
\left[\Delta\vec{p}_{\rm V\pm}({\vec v})+\Delta\vec{p}_{\rm v\pm}({\vec v})\right]\ .
\ee
The differing (anti-)symmetry properties imply that 
$\int d^3 \vec{v}\, |v_z|\,f_0^\pm(\vec{v}) \Delta\vec{p}_{4\mp}^{\ (2)}=0$.
Adding these zero terms to the non-zero terms brings Eq.~(\ref{momentum-transfer-per-area}) to a more explicit form
\be
\label{momentum-transfer-per-area-explicit}
\frac{d}{dt}\frac{d\vec{P}}{d^2\vec x_0} 
=\frac{1}{2}\,\rho_{\rm qdm}\,
\left\langle|v_z|\,\left(\frac{q}{m\,c\,v}\right)^2\,
\left(|\vec\Phi(\hat{v})|^2\,(\vec{V}-\vec{v})-(\vec{V}-\vec{v})\cdot\vec\Phi(\hat{v})\,\vec\Phi(\hat{v})
+\frac{1}{2}(\vec{V}-\vec{v})\times\vec{\Phi}_2(\hat{v})
\right)\right\rangle~,
\ee
where $\langle\cdots\rangle$ denotes the qDM particle weighted average:
\be
\langle F(\vec{v})\rangle\equiv \frac{m}{\rho_{\rm qdm}}\,
\int d^3 \vec{v}\, \,f_0(\vec{v})\,F(\vec{v})\ .
\ee
However, we note that Eq.~(\ref{momentum-transfer-per-area}) may be more useful than Eq.~(\ref{momentum-transfer-per-area-explicit}) since half the terms in a fully expanded
Eq.~(\ref{momentum-transfer-per-area-explicit}) are always zero and half the remaining terms are zero if the distribution function is $\vec{v}\to-\vec{v}$ symmetric.

This force on the qDM particles will have an equal but opposite back reaction on the ISM through the $\vec{J}_{\rm qdm}\times\vec{B}$ force.  Assuming that the entire ISM gas, with surface mass density $\Sigma_{\rm ism}$, shares this momentum, then the qDM induced ISM acceleration becomes
\be
\label{ISMacceleration}
\dot{\vec{V}}_{\rm q} = -\frac{1}{\Sigma_{\rm ism}}\frac{d}{dt} \frac{d\vec{P}}{d^2 \vec{x}_0}\ .
\ee
which must be added to the gravitational acceleration from the galaxy as a whole and to other astrophysical forces acting on the ISM.  
\subsubsection{Spin Down and other Disk Disruptions}

Here we work in the galaxy rest frame and assume the ISM follows nearly circular orbits around galactic centers.  Define the  {\it spin-down rate} for a small patch of the ISM disk as the inverse of the fractional rate of loss of ISM angular momentum, $L$, is given by
\be
\label{local-spin-down-rate}
\frac{1}{\tau_{\rm sd}}&\equiv&-\frac{1}{L}\,\frac{d}{dt}L =-\frac{\hat{V}\cdot\dot{\vec{V}}_{\rm q}}{V} ,
\ee
and we can combine Eqs.~(\ref{ISMacceleration}) and (\ref{momentum-transfer-per-area-explicit}) to write the spin down timescale as
\be
\label{teff}
\frac{1}{\tau_{\rm sd}} \equiv \frac{\rho_{\rm qdm}}{2\,\Sigma_{\rm ism}}\,
\Biggl[
\left\langle|v_z|\,\left(\frac{q\,\vec \Phi(\hat{v})\,{\rm sin}\,\psi(\hat{v})}{m\,c\,v}\right)^2\right\rangle
-\frac{\hat{V}}{V}\cdot\left\langle
|v_z|\,\left(\frac{q}{m\,c\,v}\right)^2\,\left(|\vec\Phi(\hat{v})|^2\,\vec{v}-\vec{v}\cdot\vec\Phi(\hat{v})\,\vec\Phi(\hat{v})
+\frac{1}{2}\,\vec{v}\times\vec{\Phi}_2(\hat{v})\right)
\right\rangle 
\Biggr],~~
\ee
where  $V\equiv|\vec{V}|$, $\hat{V}\equiv\vec{V}/V$, $\Phi\equiv|\vec{\Phi}|$, $\hat{\Phi}\equiv\vec{\Phi}/\Phi$ and
$\psi\equiv {\rm cos}^{-1}\hat{V}\cdot\hat{\Phi}$  is the {\it pitch angle}.  The first term comes from $\Delta\vec{p}_{\rm V+}$ and is always positive, which contributes systematically to spin-down, so we call it the {\it spin-down term}.  The second term comes from $\Delta\vec{p}_{\rm v+}+\Delta\vec{p}_{\rm v-}$, which can have either sign and will not systematically contribute to spin-down unless either the qDM velocities or magnetic field twist is correlated with rotation velocity. Note that neither of the surviving terms shown in Eq.~(\ref{teff}) depends
on $\Delta\vec{p}_{\rm V-}$, which contributes zero. We expect that, after averaging over time and different locations in the disk, that the spin-down term will dominate the spin-down rate in a realistic treatment of this scenario, but a full simulation is beyond the scope of this work and warrants further study. In Sec. \ref{IdealizedGalaxy} below, we show that, in an idealized galaxy model, this is, indeed, the only non-zero contribution to ISM spin-down.

All the terms in \Eq{teff} transfer momentum and energy between the qDM halo and ISM disk. This transfer disrupts orderly circular orbits and this disorder  eventually dissipates through radiative cooling, wind generation, or by ejecting ISM gas from the disk into the halo.  If the ISM cooling timescale is much less than the spin down timescale and the mass/energy loss to winds is small then elements of the ISM will remain in nearly circular orbits throughout the process of momentum transfer, but their radii $R$ will shrink over time.  If this occurs in the flat rotation curve part of a galactic halo, characterized by circular velocity $V_{\rm circ}$, then angular momentum conservation dictates that
\be
\label{shrink-rate}
\frac{\dot{R}}{R}=-\frac{1}{\tau_{\rm sd}}~,~
\ee
while energy conservation requires cooling with luminosity per unit area
\be
\label{spin-down-luminosity}
\frac{d{\cal L}}{d^2 \vec{x}_0}=-\Sigma_{\rm ism}\,\dot{\vec{V}}_{\rm sd}\cdot\vec{V}
=\frac{\rho_{\rm qdm}}{2\,\tau_{\rm sd}}\,V_{\rm circ}^2~,~
\ee
(the Virial theorem requires decreased potential energy equal to twice the energy loss and increased kinetic/thermal energy equal to minus the energy loss).  Joule heating of the ISM may also occur as qDM will short out the electric fields of Eq.~(\ref{MHD approximation}) in the qDM frame which will lead to generation of electric fields in the ISM frame which will, in turn, be shorted by the highly conducting ISM medium.  The long term effect of the decreased radius will likely be more noticeable than the immediate effect of radiative cooling of the excess energy.    These disruptions of the ISM disk by qDM bears further study but are outside the scope of this paper; below we set limits on qDM on the assumption that total ISM spin-down over would lead to observable consequences
of the disk lifetime. 

\subsection{Idealized Galaxy}
\label{IdealizedGalaxy}
Up until this point the derivation has been general but now we begin to make reasonable simplifying assumptions in order to derive specific bounds on qDM. 

\subsubsection{Planar Magnetic Fields}

If we assume that $\vec B$ only varies slowly in directions parallel to the plane of the disk, meaning over length scales much larger than the disk thickness, then one may approximate the magnetic field as having planar symmetry: $\vec{B}(\vec{x},z)\to\vec{B}(z)$ since most qDM trajectories, $\vec{v}$, will pass through the same vertical field structure.  For planar fields the pitch angle is $\hat{v}$ independent,  $\psi(\hat{v})\to\psi$, and our ``voltages" become
\be
\vec{\Phi}(\hat{v})   \to \frac{      \vec{\Phi}(\hat{z})    }{   |\hat{z}\cdot\hat{v} | }~~~,~~~
\vec{\Phi}_2(\hat{v})  \to  \frac{   \vec{\Phi}_2(\hat{z})\,{\rm sgn}(\hat{z}\cdot\hat{v}) }{ (\hat{z}\cdot\hat{v})^2} , 
\ee
 where the geometrical factor, $1/|\hat{z}\cdot\hat{v}|$, reflects the longer trajectory of lower inclination orbits through the disk, which increase the deflection.  Note that, from the definition of $\vec \Phi(\hat u)$ in  \Eq{defPhi}, this substitution is tantamount to the relation
 \be
 \vec \Phi(\hat v) \equiv \int_{-\infty}^{\infty} du\, \vec B(\vec x_0 + \hat v u) 
 ~~\to~~ \frac{1}{|\hat z \cdot \hat v|} \int_{-\infty}^{\infty} du\,\vec B(\vec x_0 + \hat z u) =  \frac{  \vec \Phi(\hat z) }{|\hat z \cdot \hat v|}~~,
 \ee 
 where, in this planar $\vec B$ limit, the integral can be performed along the $\hat z$ direction and the  geometric factor accounts for the path length difference of this choice. With these approximations, Eq.~(\ref{teff}) becomes 
\be
\frac{1}{\tau_{\rm sd}}=
\frac{\rho_{\rm qdm}}{2\Sigma_{\rm ism}}\,\left(\frac{q}{m\,c}\right)^2\,
\left[ 
\Phi(\hat{z})^2\,
\left({\rm sin}^2\psi\,\left\langle\frac{1}{|v_z|}\,\right\rangle
-\left\langle\frac{\hat{V}\cdot\vec{v}-\vec{v}\cdot\hat\Phi(\hat{z})\,{\rm cos}\,\psi}{V\,|v_z|}\right\rangle\right)
-\frac{1}{2}\,\frac{\hat{V}}{V}\cdot\left(\left\langle\frac{\vec{v}}{v_z}\,\right\rangle\times\vec{\Phi}_2(\hat{z})\right)~
\right]~,~~
\ee
where we have used $v_z=v\,|\hat z \cdot \hat v| $.  Note that the $v_z$ dependence of the denominator is problematic as it will lead to a logarithmic divergence in Eq.~(\ref{momentum-transfer-per-area}), which one must regulate; we address this issue below.

Since $\vec{\nabla}\cdot\vec{B}=0$ the planar approximation implies that $B_z\equiv\hat{z}\cdot\vec{B}$ must be a constant.  It is believed that the dominant ordered field is parallel to the disks, so in addition to the assumption of planar symmetry we also assume that $B_z=0$. Furthermore a planar nonzero $B_z$ necessarily extends into the halo which we have already assumed has negligible $\vec{B}$.

Planar symmetry is a convenient idealization but a poor description of the actual $\vec{B}$ in the local disk and presumably in other parts of our Galaxy and in other galaxies as well.  Unfortunately it is difficult to determine the full field geometry observationally.  Even locally some components of $\vec{B}$ are much more uncertain than others.  Given these uncertainties we proceed assuming $\vec{B}$ is planar and $B_z=0$.

\subsubsection{Momentum Transfer Saturation}
\label{saturation}

The $v_z\to 0$ limit of the 2nd Born approximation is not accurate because the path length of low inclination orbits  particles is so long that their deflection becomes large invalidating the approximation and overestimating the momentum transfer. One expects the $v_z \to 0$ trajectories to already be co-rotating with the ISM and $\Delta\vec{p}\to0$ so the assumption that the distribution is not significantly modified is in error.  However since $\langle|v_z|^{-1}\rangle$ is only logarithmically divergent for smooth $f_0(\vec{v})$  it does not make too much difference how one regulates it.  We take the simple approach of modifying the momentum transfer rather than the distribution function.  In particular, we choose the replacement
\be
\label{saturation}
\left\langle\frac{1}{|v_z|}\,\right\rangle\to\left\langle{\rm min}\left(\frac{1}{|v_z|},\frac{|v_z|}{v_*^2}\right)\right\rangle~~~,~~
v_*\equiv\sqrt{2}\,\frac{q\,\Phi(\hat{z})\,{\rm sin}\psi}{m\,c}~,
\ee
where $v_*$ is roughly the value of $|v_z|$ where large deflections occur in a single disk crossing, i.e. $v_*\sim v\,\Delta z/R_{\rm L}$.  Note that $\langle\vec{v}/|\hat{v}|\rangle$ in the twist term is not divergent and needs no regulation so we do not modify it even though a modified distribution $f_0$ would change it slightly.  While Eq.~(\ref{saturation}) does remove the mild divergence we emphasize that this cutoff could be implemented more rigorously, but doing so is beyond the scope of this work.  

\subsubsection{Maxwellian Velocity Distribution}

For simplicity, we adopt a  Maxwellian velocity distribution in the galaxy frame
\be
f_0(\vec v) = \frac{ \rho_{\rm qdm}}{(2\pi)^{3/2} m\sigma^3}\,e^{-\frac{|\vec{v}|^2}{2\sigma^2}}\ .
\ee
where $\sigma$ is the 1-D velocity dispersion.  Thus we are assuming the qDM halo is not rotating. This is a symmetric distribution, $f_0^-=0$, so $\Delta\vec{p}_{\rm V-}$ and $\Delta\vec{p}_{\rm v-}$ are irrelevant.  The velocity averages are
\be
\left\langle{\rm min}\left(\frac{1}{|v_z|},\frac{|v_z|}{v_*^2}\right)\right\rangle= 
\frac{1}{\sqrt{2\pi\sigma}}\left(\Gamma(0, \varepsilon^2) +\frac{1-e^{-\varepsilon^2}}{\varepsilon^2}\right)~~,~~~
\left\langle\frac{\vec{v}}{|v_z|}\right\rangle=\vec{0} ~~~,~~~
\left\langle\frac{\vec{v}}{v_z}\right\rangle=\,\frac{1}{2}\,\hat{z}
\ee
where 
$
\Gamma(s,x) \equiv \int_x^\infty dt \, t^{s-1} e^{-t}
$
is the upper incomplete Gamma function and we have defined the parameter
\be
\varepsilon\equiv\frac{v_*}{\sqrt{2}\sigma}=\frac{q\,\Phi(\hat{z})\,{\rm sin}\,\psi}{m\,c\,\sigma}\ .
\ee
Note that $\varepsilon\sim\Delta z/R_{L} \ll1$.   Since we are working in the ballistic limit, $\varepsilon\ll1$, we use the asymptotic expression
$\Gamma(0, \varepsilon^2)+\varepsilon^{-2}\,(1-e^{-\varepsilon^2})\to1-\gamma_{\rm E}-2\,{\rm ln}\,\varepsilon$, where $\gamma_{\rm E} \ = 0.577216...$ is Euler's constant.  With this substitution, the local spin-down time is approximately
\be
\frac{1}{\tau_{\rm sd}}\approx
\frac{\sigma\,\rho_{\rm qdm}}{2\,\Sigma_{\rm ism}}\,\,
\left[\left(\frac{q\,\Phi(\hat{z})\,{\rm sin}\,\psi}{m\,c\,\sigma}\right)^2\,
\frac{1-\gamma_{\rm E}-2\,{\rm ln}\,\frac{q\,\Phi(\hat{z})\,{\rm sin}\,\psi}{m\,c\,\sigma}}{\sqrt{2\pi}}
+\frac{\sigma}{4V}  \left(\frac{q}{m\,c\,\sigma}\right)^2\,
\left(\hat{z}\times\hat{V}\right)\cdot\vec{\Phi}_2(\hat{z})
\right]\ .
\ee
If one defines $\hat{z}$ to be in the direction of the angular momentum vector of galactic rotation (rather than the opposite direction), then $\hat{z}\times\hat{V}=-\hat{r}$ where $\hat{r}$ is the direction away from the galaxy center.  With planar symmetry this implies 
\be
\hat{r}\cdot\vec{\Phi}_2(\hat{z})=\int dz \int dz'\,{\rm sgn}(z-z')\,\left(  B_\theta(z)\,B_z-B_z\,B_\theta(z') \right),
\ee
where $B_\theta(z)\equiv\hat{V}\cdot\vec{B}(z)$.   This is an ill-defined integral if $z\to\pm\infty$, but vanishes due to $z\leftrightarrow z'$ exchange symmetry for any finite $z$ interval.  Furthermore under the assumption of fields in the plane, $B_z=0$, the integral is again zero.  Under our assumptions of planar fields and a non-rotating halo (the specific Maxwellian form of the distribution does not matter here) twisted fields do not contribute to spin-down.  Thus the final expression for the local spin-down rate under the 2nd Born and other approximations is
\be
\label{spin-down-rate-final}
\frac{1}{\tau_{\rm sd}}\approx
\frac{\sigma\,\rho_{\rm qdm}}{2\,\Sigma_{\rm ism}}\,\left(\frac{q\,\Phi(\hat{z})\,{\rm sin}\,\psi}{m\,c\,\sigma}\right)^2\,\frac{1}{\sqrt{2\pi}}
\left[         1-\gamma_{\rm E}-2\,{\rm ln} \left( \frac{q\,\Phi(\hat{z})\,{\rm sin}\,\psi}{m\,c\,\sigma} \right) \right] \ ,
\ee
where  only the spin-down term contributes.

\subsubsection{Mean Spin Down Rate}

We have computed a local spin-down rate in terms of $\tau_{\rm sd}(\vec{x}_0,t)$.  We fully expect $\tau_{\rm sd}$ to fluctuate within a galaxy and over the history of a galaxy.  In the flat rotation curve and rapid cooling regime the radius of the ISM will decrease according to \Eq{shrink-rate}, whose formal solution has the exponential form 
\be
R(t)\propto {\rm exp}\left(-\int^t \frac{dt'}{\tau_{\rm sd}(t')}\right)~,~~
\ee
so $\tau_{\rm sd}$ should not significantly exceed the age of the galaxy in order to avoid catastrophic spin-down.  Henceforth, we will speak of typical values of for the spin-down rate, $\bar{\tau}_{\rm sd}$ though we do not perform any specific averaging.

The numerical value of $\bar{\tau}_{\rm sd}$ is the quantity we will use to set limits on qDM particles.  Note that the only quantities in this expression that depend on the particle properties of qDM is the ratio $q/m$ and $\rho_{\rm qdm}$.  The other quantities are, in principle, measurable properties of the ISM and DM halo.  While we may not know the mass density of qDM in a galaxy, we do have dynamical information on the total density of dark matter $\rho_{\rm dm}$.  If we parameterize $\rho_{\rm qdm}=f_{\rm qdm}\,\rho_{\rm dm}$ with $0\le f_{\rm qdm}\le1$ then we see that ISM spin-down sets limits on the combination $\sqrt{f_{\rm qdm}}\,q/m$ but with one important caveat discussed below.

\subsubsection{qDM Spin Up}

The caveat is that if $f_{\rm qdm}$ is too small then the inertia of the qDM is not sufficient to significantly spin-down the ISM; rather the qDM will be {\it spun-up} to co-rotate with the ISM without significantly affecting the ISM.  A rough estimate of when spin-up can occur is when $R\,\rho_{\rm qdm}\le\Sigma_{\rm ism}$.  This condition may be written $f_{\rm qdm}\le f^*_{\rm qdm}\equiv\Sigma_{\rm ism}/(2\,R\,\rho_{\rm dm})$.  For these small values of $f_{\rm qdm}$, ISM spin-down does not constrain qDM parameters. However, this effect can alter the qDM velocity distribution, which may have nontrivial implications for
experimental detection; a proper numerical simulation beyond the scope of this work is required to properly assess the importance of this effect.

While qDM spin-up may not pose any obvious observational constraints it could be of interest if this occurred since a spun-up qDM velocity distribution function would differ significantly from the normal expectation and in particular one would expect no qDM wind in the disk rest frame.  In analogy with the spin-down rate one can define a spin-up rate by rescaling by the relative inertial mass of the qDM and the ISM:
\be
\label{SpinUpRate}
\frac{1}{\tau_{\rm su}}\approx\frac{\Sigma_{\rm ism}}{2 R \rho_{\rm qdm}}\,\frac{1}{\tau_{\rm sd}}
=\frac{1}{4}\,\frac{\sigma}{R}\,
\left(\frac{q\,\Phi(\hat{z})\,{\rm sin}\psi}{m\,c\,\sigma}\right)^2\,
\frac{1}{\sqrt{2\pi}}
\left[         1-\gamma_{\rm E}-2\,{\rm ln} \left( \frac{q\,\Phi(\hat{z})\,{\rm sin}\,\psi}{m\,c\,\sigma} \right) \right] \ ,
\ee
which is independent of $\rho_{\rm qdm}$ and $\Sigma_{\rm ism}$ because in this limit the qDM is merely a spectator particle to the magnetic fields.  The quantity $R/\sigma$ is roughly the dynamical timescale of the dark matter.  The condition for spin-up is that $\bar{\tau}_{\rm su}$ is significantly shorter than the age of the disk.

\subsubsection{Spin Down Limits on Charged Dark Matter}

While it is fully expected that the magnetic field ($\Phi(\hat{z})$) will increase over time as the a galactic dynamo grows the field, the field direction, $\hat{V}\cdot\hat{\Phi}$, will vary, and the amount of ISM gas ($\Sigma_{\rm ism}$) will decrease over time as gas is turned into stars (although infalling gas will replace some of it).  Here we  suppose that these evolutionary effects can be modeled as if the current configuration has been constant over an interval of duration $T_{\rm ism}$. 

To set our approximate limit the requirement that an ISM disk not be significantly affected by spin-down is 
$\bar{\tau}_{\rm sd}>T_{\rm ism}$, or equivalently
\be
\frac{1}{\sqrt{f_{\rm qdm}}}\,\frac{m\,c^2}{q}>
 \Phi_{\rm eff}\,
\frac{c}{\sigma}\,
\sqrt{\frac{\rho_{\rm dm}\,\sigma\,T_{\rm ism}}{2\sqrt{2\pi}\,\Sigma_{\rm ism}}}\,
\sqrt{1-\gamma_{\rm E} 
-2\, {\rm ln}\left(\frac{q\,\Phi_{\rm eff}}{m\,c^2}\,\frac{c}{\sigma}\right)}
~~~,~~~\Phi_{\rm eff}\equiv\Phi(\hat{z})\,|{\rm sin}\psi|\ .
\ee
The last term is a logarithmic correction of order unity,
$\rho_{\rm dm}\,T_{\rm ism}\,\sigma/(\Sigma_{\rm ism})$ is a large dimensionless number approximately equal to the ratio of masses of dark matter and ISM times the age of the disk in units of its dynamical timescale, and $c/\sigma$ is another large dimensionless number.  These large factors multiply the already very large voltage $\Phi_{\rm eff}$.

While one cannot express this limit analytically due to the logarithm, a first order iterative solution is fairly accurate:
\be
\label{iterative-spin-down-bound}
\frac{1}{\sqrt{f_{\rm qdm}}}\,\frac{m\,c^2}{q}
>\Phi_{\rm eff}\,
\frac{c}{\sigma}\,
\sqrt{\frac{\rho_{\rm dm}\,\sigma\,T_{\rm ism}}{2\,\sqrt{2\pi}\,\Sigma_{\rm ism}}}\,
\sqrt{1-\gamma_{\rm E} 
+{\rm ln}\left(f_{\rm qdm}\,\frac{\rho_{\rm dm}\,T_{\rm ism}\,\sigma}{2\,\sqrt{2\pi}\,\Sigma_{\rm ism}}\right)}\ .
\ee
Similarly, one can compare this to the condition for qDM spin-up $\bar{\tau}_{\rm su}<T_{\rm ism}$, which in the same iterative approximation yields the condition
\be
\label{iterative-spin-up-bound}
\frac{m\,c^2}{q}
<\Phi_{\rm eff}\,
\frac{c}{\sigma}\,
\sqrt{\frac{\sigma\,T_{\rm ism}}{4\sqrt{2\pi}\,R}}\,
\sqrt{1-\gamma_{\rm E}+{\rm ln}\left(\frac{\sigma\,T_{\rm ism}}{4\sqrt{2\pi}\,R}\right)}\ .
\ee
However, recall that spin-up is not excluded, but only indicates a major modification of the qDM velocity distribution.

\section{The Local Galaxy}
\label{galaxy}
\subsubsection{Global Galactic Parameters}

As mentioned previously, we will first apply qDM limits on spin-down by looking at the Galactic disk in the region of the solar neighborhood; not because it is the optimal location from which to derive the tightest constraints, but rather because we have better information of the properties of the nearby disk.  That being said, there remain order unity uncertainties in the inventory of material in local Galaxy, although we do know the solar radius $R=8.2\pm0.1\,{\rm kpc}$ and solar rotation velocity $V=248\pm3\,{\rm km/s}$ fairly accurately \cite{BlandHawthorn2016}.

The local halo dark matter properties can be estimated by the large scale gravitational field of the Galaxy which is traced by stellar orbits, although there remain uncertainties associated with the unseen and possibly anisotropic velocity dispersion of the dark matter as well as uncertainties in the baryonic content.  Here we use canonical values $\rho_{\rm dm} = 0.3\,{\rm GeV/cm^3}/c^2=0.008\,M_\odot/{\rm pc}^3$ \cite{Bovy:2012tw}, and $\sigma = 270 \, {\rm km/s}$ \cite{Jungman1996}.

From local observations one cannot determine the history of the magnetic field in the Solar or any other neighborhood to inform a choice for $T_{\rm ism}$.   While there is little data on magnetic fields in galaxies that are not near to us ($z\ll1$)  Mao et al. \cite{Mao2017} has measured the rotation measure (RM) of quasar light passing through a galactic disk at a time $\sim5\,{\rm Gyr}$ ago ($z=0.439$); finding RM and inferred $B$ values similar to that seen nearby galaxies.  MHD simulations of mock galaxies meant to be similar to the Milky Way by Pakmor et al. \cite{Pakmor2017} show a variety of time histories but in many cases the magnetic field strengths were stable or decreasing for the past $\sim5\,{\rm Gyr}$.  Motivated by these measurements and simulations we use $T_{\rm ism}\sim5\,{\rm Gyr}$.

\subsubsection{Inventory of the Disk Contents}

The inventory of gas and stars is also uncertain since the disk extends to quite a large distance above and below us, and much of this mass is difficult to detect and localize at large distance, e.g. brown dwarfs and atomic hydrogen.  Different methods have been employed to estimate the inventory such as  starting with the density and velocity dispersions of very local matter and then self-consistently solving for the gravitational potential to extrapolate to the full thickness of the disk.  Table 2 of Ref.~\cite{Read:2014qva} estimates the baryonic content of the disk as $3.0\pm1.5,\,12.0\pm4.0,\,2.0\pm1\,M_\odot/{\rm pc^2}$ for molecular, atomic and ionized gas, respectively and $30.0\pm1,\,7.2\pm0.7\,M_\odot/{\rm pc^2}$ for burning and non-burning stars.  From this, the total baryonic surface density would be $\Sigma_{\rm disk}=54.2\pm4.9\,M_\odot/{\rm pc^2}$.  For spin-down only the gaseous material matters which has a surface density of $\Sigma_{\rm ism}=17.2\pm4.4\,M_\odot/{\rm pc^2}$.  Estimates such as these make assumptions with associated uncertainties not represented in the quoted errors so there probably are additional uncertainties.

One might also consider the existence of an additional {\it dark disk} component \cite{Purcell2009,Fan2013} which has been limited to $\Sigma_{\rm dd}<14\,M_\odot/{\rm pc^2}$ if its scale height is $\le100\,$pc \cite{Schutz:2017tfp}. Such a dark disk would not be counted as part of the ISM, but its gravity would result in lowering of the estimates of the mass of other components including the ISM.  Such a dark disk might contain a small part of the total dark matter and, since these orbit inside of the ISM, would presumably be co-rotating with the stars and gas.  Any dark disk qDM would not play a dominant role in spin-down since its inertial mass is not large compared to that of the ISM.  One does not expect spin up of halo qDM to contribute to condensation of a dark disk from the halo since any dark disk has much larger phase space density than the rest of the DM requiring a dissipation mechanism not provided by interactions with $\vec{B}$.

\subsubsection{Local Magnetic Field}

A simple picture of magnetic fields in spiral galaxies is that $\vec{B}$ lies in the plane of the galaxy along the spiral arms.  In this case, the pitch angle, $\psi$, is the angle between the spiral arms and the circular orbits.  If the spiral arms are wound tightly, then $\psi$ could be quite small, which would weaken the spin-down constraints.  In a coherent planar magnetic field model, where the $\vec{B}$ and $n_{\rm e}$ (free electron density) have exponential scale heights $h_{B}$ and $h_{\rm e}$ respectively, the disk contribution to the so-called rotation measure (RM) to extra-Galactic radio sources should be given by a dipole pattern
\be
\label{RM}
{\rm RM}({\hat n})
=\frac{-e^3}{2\pi\,(m_{\rm e}\,c^2)^2}\int_0^\infty dr\,n_{\rm e}\,\hat{n}\cdot\vec{B}
=\frac{-e^3\,N_{\rm e}}{4\pi\,(m_{\rm e}\,c^2)^2}\,\frac{h_{B}}{h_{B}+h_{\rm e}}
\frac{{\hat n}\cdot\vec{\bar{B}}_0}{|\hat{n}\cdot\hat{z}|}~,
\ee
where $\vec{\bar{B}}_0$ gives the field in the plane, $N_{\rm e}$ the vertical column density of free-electrons and $|\hat{n}\cdot\hat{z}|$ accounts for the increased electron column for lines-of-sight not perpendicular to the disk.  One can estimate $N_{\rm e}$ from dispersion measure (DM), defined
according to 
\be
{\rm DM}=\int_0^\infty dr\,n_{\rm e}~,
\ee
 and the YMW16 \cite{Yao2017} model of the Galactic electron gives ${\rm DM}(\pm\hat{z})=18.99\pm0.57\,{\rm pc/cm^3}$ summing to 
$N_{\rm e}=\sum_\pm{\rm DM}(\pm\hat{z})=1.17\times10^{20}{\rm cm^{-2}}$.  Here we adopt a round number for the exponential scale heights $h_{B}= h_{\rm e}=1\,{\rm kpc}$ \cite{Sun2007} although these numbers are not known accurately.

Measured RM maps are significantly more complicated than a dipole pattern.  Decomposing RM maps derived from extra-Galactic  radio sources (\cite{Oppermann:2011td}) into reflection even and odd parts, 
${\rm RM}_\pm(\hat{n})=\frac{1}{2}({\rm RM}(+\hat{n})\pm{\rm RM}(-\hat{n}))$,
one finds $\langle|\hat{n}\cdot\hat{z}|\,{\rm RM}_+\rangle_{\rm rms}=14\,{\rm rad/m^2}$ and
$\langle|\hat{n}\cdot\hat{z}|\,{\rm RM}_-\rangle_{\rm rms}=11\,{\rm rad/m^2}$; nearly equal contributions.  It is only the ${\rm RM}_-$ component which gives evidence for a non-zero line integral of $\hat{v}\cdot\vec{B}$, which could contribute a non-zero component to $\vec{\Phi}$.  The ${\rm RM}_+$ contribution could indicate a field reversal above and below the Galactic plane \cite{Sun2007,Han2017}.  If $\vec{B}$ were perfectly antisymmetric,  then $\vec{\Phi}=0$ and the spin-down term would be zero!  If in addition the free electron density were symmetric about the plane then ${\rm RM}_-=0$.  Since even ${\rm RM}_+$ and ${\rm RM}_-$ are nearly the same in magnitude it seems likely $\vec{B}$ is far from perfectly antisymmetric about the plane.

One can invert Eq.~(\ref{RM}) to estimate the dipole component of RM:
\be
\label{RMinv}
\vec{\bar{B}}_0
=-3\frac{(m_{\rm e}\,c^2)^2}{e^3}\,\frac{h_{B}+h_{\rm e}}{h_{B}}\,
\frac{\int d^2\hat{n}\,|\hat{n}\cdot\hat{z}|\,{\rm RM}({\hat n})}{N_{\rm e}}
\approx0.76\,\mu{\rm G}\,\hat{b} ~~,~~ \hat{b}=(\ell\approx89^\circ,b\approx15^\circ) ~~,~~ \psi\approx15^\circ\ ,
\ee
where we have used maps from \cite{Oppermann:2011td} for these numerical values.  This $|\vec{\bar{B}}_0|$ is a few times larger than  previous estimates of $\sim0.2\,\mu{\rm G}$ coherent fields \cite{Brown2007,Brown:2010jx}.  To set a central limit on qDM below, we use Eq.~(\ref{RMinv})  which gives the estimate
$\Phi_{\rm eff}=2\,h_{B}\,|\vec{\bar{B}}_0\,{\rm sin\psi}|=3.5\times10^{17}\,$Volts.

The dipole contributes very little to the ${\rm RM}_-$ pattern:
$\langle|\hat{n}\cdot\hat{z}|\,{\rm RM}_-\rangle^{\rm dipole}_{\rm rms}=3.4\,{\rm rad/m^2}$, indicating that the coherent component of $\vec{B}$ is a small part of the overall field pattern, and (being nearly parallel to $\vec{V}$) applies little torque relative to the field strength; most of the $\vec{J}_{\rm qdm}\times{B}$ force is radial.  While there is no definite relationship between ${\rm RM}_-(\vec{n})$ and $\vec{\Phi}(\hat{n})$, and it is even possible that the former is large when the later is zero, a larger overall ${\rm RM}_-$ is suggestive of large torque contributions from higher multipole fields.  There are many hot and cold regions in the ${\rm RM}$ maps associated with Galactic features that are not balanced by similar features in the opposite direction.  Using the rms ${\rm RM}_-$ rather than the dipole rms would increase the field strength by a factor of $\sim 4$ and the larger pitch angle for other lines of sight could increased $\langle|\sin\psi|\rangle_{\rm rms}$ by $\sim3$ leading to
${\Phi}_{\rm eff}\sim5\times10^{18}\,{\rm Volts}$.  
While we have largely dismissed the notion of perfectly antisymmetric $\vec{B}$ fields in the Galaxy, we should note that even anti-symmetric fields do not predict zero spin-down except under the incorrect assumption of planar symmetry.  There are radial and azimuthal non-planar gradients in the fields which will contribute a non-zero $\Phi_{\rm eff}$ even with perfect antisymmetry.  The gradients are on the $R\sim8\,{\rm kpc}\sim8\times h_{B}$  scale, which would only suppress $\Phi_{\rm eff}$ by about factor of $\sim0.1$.

In summary, there are large uncertainties on the numerical value of $\Phi_{\rm eff}$, at least an order of magnitude.  Reducing the uncertainty of this local quantity would require more precise and accurate modeling of $\vec{B}$.  However, in other parts of our Galaxy (or in other galaxies), these uncertainties are even larger, so in what follows, we use the local Galaxy numbers extracted in this section to place limits on qDM.

\begin{figure*}
\hspace{-0.2in}
\includegraphics[width=4 in,angle=0]{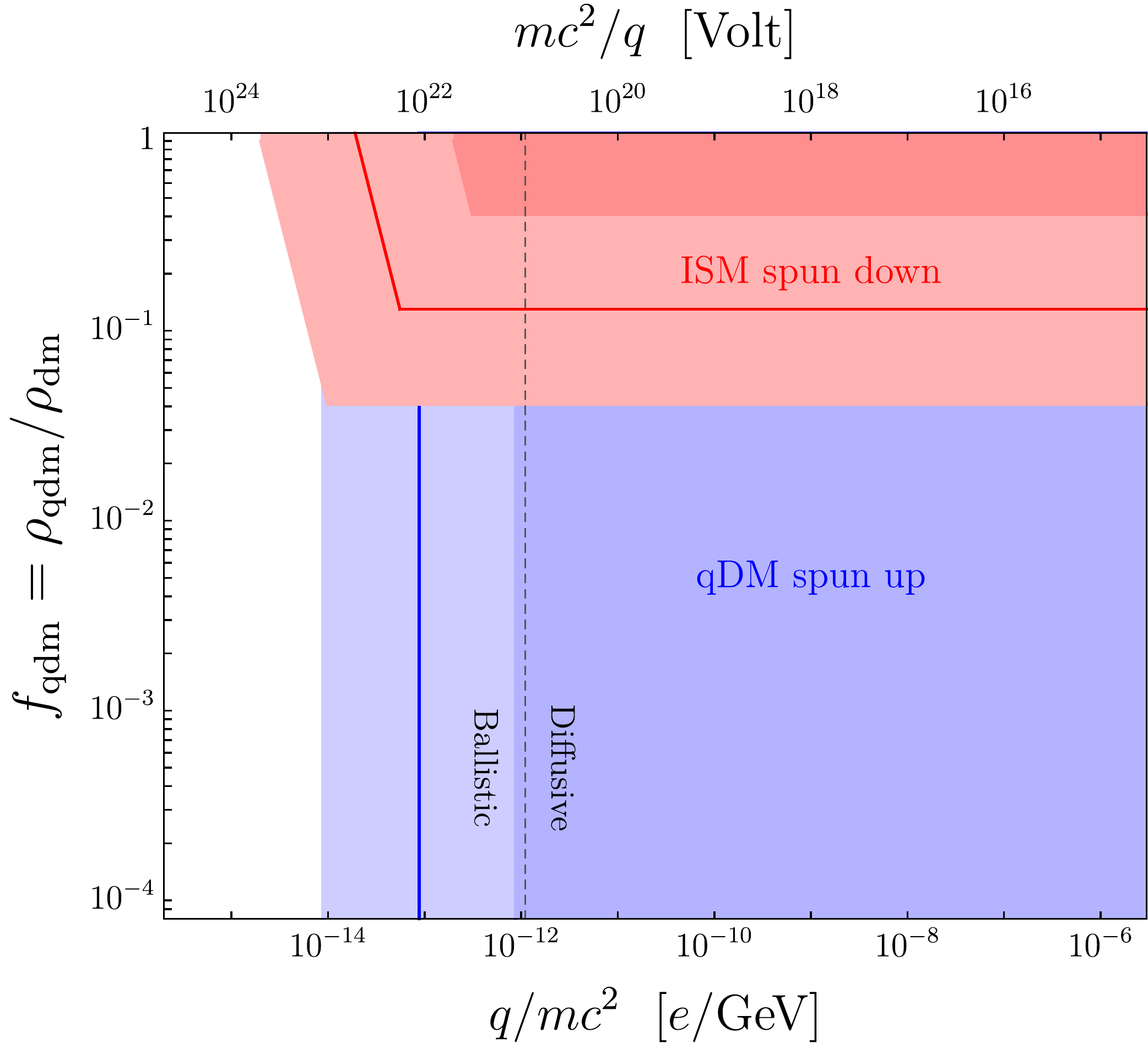}~~~~
\caption{Shown are our expectation for limits on charged dark matter (qDM) parameters from {\it spin-down} of the interstellar medium ISM (see text) in regions of spiral galaxy disks similar to the Solar neighborhood.  The dark pink region we believe is almost certainly ruled out while the lighter pink region might be ruled out.  The central value of the bound on qDM is the red line.  The upper and lower horizontal axes are just two parameterizations of the charge to mass ratio of qDM while the vertical axis is the ratio of mass in qDM to that of all dark matter ($f_{\rm qdm}$ in the text).  In the dark blue region the qDM almost certainly is {\it spun up}, meaning it will be nearly co-rotating with the Disk, while in the light blue region it might be; the solid blue line is our central estimate of the boundary between spun-up and non spun-up qDM halos.  The black dashed line separates parameter values where a typical qDM particle would pass through the nearby disk with only slight deflection by the disk magnetic field (the ballistic regime) and values where orbits are convoluted (the diffusive regime).  Our detailed calculation of spin-down is valid in the ballistic regime.
}
\label{qDMlimits}
\end{figure*}

\subsubsection{Local Limits on qDM}

We have motivated the local values $\rho_{\rm dm}=0.008\,M_\odot/{\rm pc^3}$, $\Sigma_{\rm ism}=17.2\,M_\odot/{\rm pc^2}$, $\sigma=270\,{\rm km/sec}$, $\Phi_{\rm eff}=10^{18}\,{\rm Volt}$, $\psi=15^\circ$ and $T_{\rm ism}=5\,{\rm Gyr}$.   Substituting these values into Eq.~(\ref{spin-down-rate-final}) we find the spin-down timescale to be
\be
\bar{\tau}_{\rm sd}&\approx&\frac{0.6\,{\rm Gyr}}{f_{\rm qdm}\,(1+0.2\,{\rm ln} f_{\rm qdm})}\,
\left(\frac{m\,c^2}{{\rm GeV}}\,\frac{10^{-13}}{q/e}\right)^2\ ,
\ee
so demanding that this does not exceed $T_{\rm ism}$ yields the qDM limit 
\be
\label{local-spin-down-bound}
\frac{m\,c^2}{q}>2.9\times10^{22\pm1}\,{\rm Volt}\,\sqrt{f_{\rm qdm}\,(1+0.2\,{\rm ln} f_{\rm qdm})}
\qquad {\rm or}\qquad
\frac{q}{m\,c^2}<\frac{3.5\times10^{-14\mp1}}{\sqrt{f_{\rm qdm}\,(1+0.2\,{\rm ln} f_{\rm qdm})}}\,\frac{e}{\rm GeV}\ ,
\ee
where we have included an {\it ad hoc} two orders of magnitude of uncertainty to represent the cumulative uncertainties in the local magnetic field and matter content. Using $R=8\,{\rm kpc}$ we find that these limits should only apply when
\be
f_{\rm qdm}>f_{\rm qdm}^*=0.13\times10^{\pm0.5}\ ,
\ee
below which we expect qDM spin-up rather than ISM spin-down.  Here we have again included an {\it ad hoc} uncertainty, this time one order of magnitude.   For both of these inequalities one would need to develop a concrete observational spin-down test to determine which qDM parameters are acceptable.  We do believe that $f_{\rm qdm}=1$ is well within the spin-down region.

The parameter regime where the qDM is spun-up is equally uncertain as it also suffers from the same uncertainties in the magnetic field
\label{local-spin-up-bound}
\be
\bar{\tau}_{\rm su}&\approx&5.0\,{\rm Gyr}\,
\left(\frac{m\,c^2/q}{10^{22}\,{\rm Volt}}\ = \  
       \frac{m\,c^2}{{\rm GeV}}\,\frac{10^{-13}}{q/e}\right)^2\ ,
\ee
hence qDM spin-up becomes significant when
\be
\label{local-spin-up}
\frac{m\,c^2}{q}<1.0\times10^{22\pm1}\,{\rm Volt}\
\qquad {\rm or}\qquad
\frac{q}{m\,c^2} > 1.0\times10^{-13\mp1}\,\frac{e}{\rm GeV}\ .
\ee
For our limiting value of $\tau_{\rm sd}<5\,{\rm Gyr}$ the excess radiative cooling of the ISM from Eq.~(\ref{spin-down-luminosity}) would be $\approx0.02\,L_\odot/{\rm pc}^2$, which is probably too small to detect since it could be confused with other mechanisms which heat the ISM.

The inequalities in Eqs.~(\ref{local-spin-down-bound}) and (\ref{local-spin-up})  are the main results of this paper.  These findings are summarized graphically in Fig.~\ref{qDMlimits}, which plots the qDM halo density fraction $f_{\rm qdm} = \rho_{\rm qdm}/\rho_{\rm dm}$ against the charge to mass ratio $q/m\,c^2$ on the lower horizontal axis (equivalently the ``voltage" $m\,c^2/q$ on the upper horizontal axis). For $f_{\rm qdm} \gtrsim 0.13$, there is insufficient inertial mass passing through the ISM over a $T_{\rm ism} \sim 5$ Gyr timescale, so the qDM will be spun up.  The vertical dashed curve delineates the ballistic and diffusive regimes; the Born approximation is only valid in the former.

\section{Where are the Best Bounds?}
\label{best}

Our bounds on charged dark matter are  based on the local Galaxy and are weaker than they would be if we had similarly accurate estimates of the ISM and magnetic fields in those regions further out in our Galaxy.  This is because the ratio of dark matter to ISM is larger there: the ISM falls off roughly exponentially with $R$ while the dark matter only falls off as $\sim R^{-2}$.  The outer parts of other galaxies are more easily observed and also might yield tighter bounds.  For this reason we consider our local Galaxy bound to be a conservative estimate of what one might do with this type of constraint.

We have set bounds on qDM by their interaction with ordered magnetic fields.  Such ordered fields exist in other parts of the universe besides spiral galaxies.  One might consider whether stronger bounds might be set from observations of different types of objects, like clusters of galaxies.  The strong bound in Eq.~(\ref{iterative-spin-down-bound}) leverages several key features of an ISM disk embedded in a halo of qDM particles: 
\begin{itemize}
    \item{\bf  Velocity:} The large relative velocity between DM and gas yields more DM-ISM momentum transfer
    \item{\bf Density:} The large DM to gas density ratio enables the DM to more easily deform the disk
    \item{$\mathbf{B}$}{\bf-Field:} The  large field and coherence length enable stronger DM-ISM couplings  
    \item{\bf Timescale:}  The DM-ISM momentum transfer can accumulate over long $\gg$ Gyr timescales 
\end{itemize}
The intra-cluster medium (ICM) in clusters of galaxies can have much larger field strengths over much larger length scales than spiral galaxy disks and while clusters are younger than galaxies they are not much younger.  What clusters lack is a relative velocity between the ICM and the dark matter at least for ``relaxed" clusters.  There therefore may be no momentum or energy exchange between the qDM and ICM so there is no obvious analog to spin-down.   This is less true for clusters in the process of formation or unrelaxed clusters.  

An exceptional case of an unrelaxed cluster is the Bullet Cluster which is in the beginning stages of merging where two clusters have collided leaving the merged ICM from both of them in between two still distinct galaxy and dark matter clusters which passed through each other.  If $q/m$ is large enough (diffusive rather than ballistic) then the qDM would be bound to the ICM  and would have remained with the merged ICM, only the galaxies would have continued to move apart.  If that were the case and $f_{\rm qdm}\approx1$ then the dark matter would not have been observed (by weak lensing) to be surrounding the galaxies but rather surrounding the merged ICM.  This is similar to an argument used to put limits on self-interacting dark matter from the Bullet cluster but for qDM the interaction is with the ICM mediated by the magnetic fields.  One can roughly estimate the Bullet bound in a similar fashion as Eq.~(\ref{diffusiveBound}) by requiring the Larmor radius to be larger than the size of the clusters in order that the qDM not be stuck to the ICM:
$R_{L}=m\,c\,V_{\rm bc}/(q\,B_{\rm bc})>R_{\rm bc}$ or $m\,c^2/q>(B_{\rm bc}\,R_{\rm bc})\,(c/V_{\rm bc})$.
Here $V_{\rm bc}$ is the relative speed of the cluster when they passed through each other and $R_{\rm bc}$ the size of the ICM containing the magnetic field.  We have no direct knowledge of $B_{\rm bc}$ at the time of the merger but simulations suggest (disordered) peak fields as high as $60\,\mu{\rm G}$.  Using studies of other clusters as a guide one might guess a pre-collision ordered field strength of $10\,\mu{\rm G}$ and  ICM size $R\sim 50\,{\rm kpc}$.  This yields a voltage $B_{\rm bc}\,R_{\rm bc}\sim5\times 10^{20}\,{\rm Volt}$ or $\sim10^3$ times larger than $\Phi_{\rm eff}$ obtained for the local Galactic disk.  An impact velocity of $\sim 10^3\,{\rm km/sec}$ can be inferred from observables so a rough estimate of the bound that might be obtained from the Bullet cluster is
\be
\label{BulletBound}
\frac{m\,c^2}{q}\gtrsim10^{23}\,{\rm Volt}\,
\left(\frac{B_{\rm bc}}{10\,{\rm\mu G}}\right)\,
\left(\frac{R_{\rm bc}}{50\,{\rm kpc}}\right)\,
\left(\frac{1000\,{\rm km/sec}}{V_{\rm bc}}\right) \quad {\rm when} \quad f_{\rm qdm}\approx 1\ .
\ee
This is an order of magnitude stronger than our central spin-down bound from the local Galaxy. It is not much much greater than the spin-down bound because there is no cumulative effect over Gyrs but only the effect of one passage of qDM through the ICM of the colliding cluster. This rough analysis suggests that rare cluster collisions give bounds comparable to spin-down and it is not clear which system is better able to constrain the properties of qDM.  It is likely true that clusters limits would not be able to constrain as small a value of $f_{\rm qdm}$ as spin-down since for these messy systems one could easily loose track of a fraction of the dark matter.

It was suggested in Ref.~\cite{2016arXiv160204009K} that the density profile of even relaxed clusters would be noticeably affected by qDM and an observational bound of $q/(m\,c^2)\lesssim10^{-14}\,{e/{\rm GeV}}$ might be therefore be set, which is precisely the same bound as Eq.~(\ref{BulletBound}), a factor of 10 times stronger than our central galactic spin-down bound.   As with galactic spin-down detailed simulations in comparison with observables  are required to accurately set precise qDM limits from clusters of galaxies, whether relaxed or not.




\section{Theoretical Implications: from Millicharges to WIMPzillas}
\label{models}

Armed with a specific bound on qDM properties, we can now discuss implications of our analysis for particular dark matter scenarios.
The ISM spin-down bound is particularly interesting because it probes parts of $q$-$m$ parameter space where the qDM may produced with non-thermal abundance either because of the weak coupling  ($q\ll e$ milli-charges) or because of the high mass ($m\sim10^{13}\,{\rm GeV}$ WIMPzillas \cite{Kolb:1998ki}). Detection of a non-thermal abundance of relic particle probes aspects of early universe physics that are not accessible with particles whose abundance is determined by simple thermal equilibrium.

\begin{figure*}
\hspace{-0.2in}
\includegraphics[width=4.5 in,angle=0]{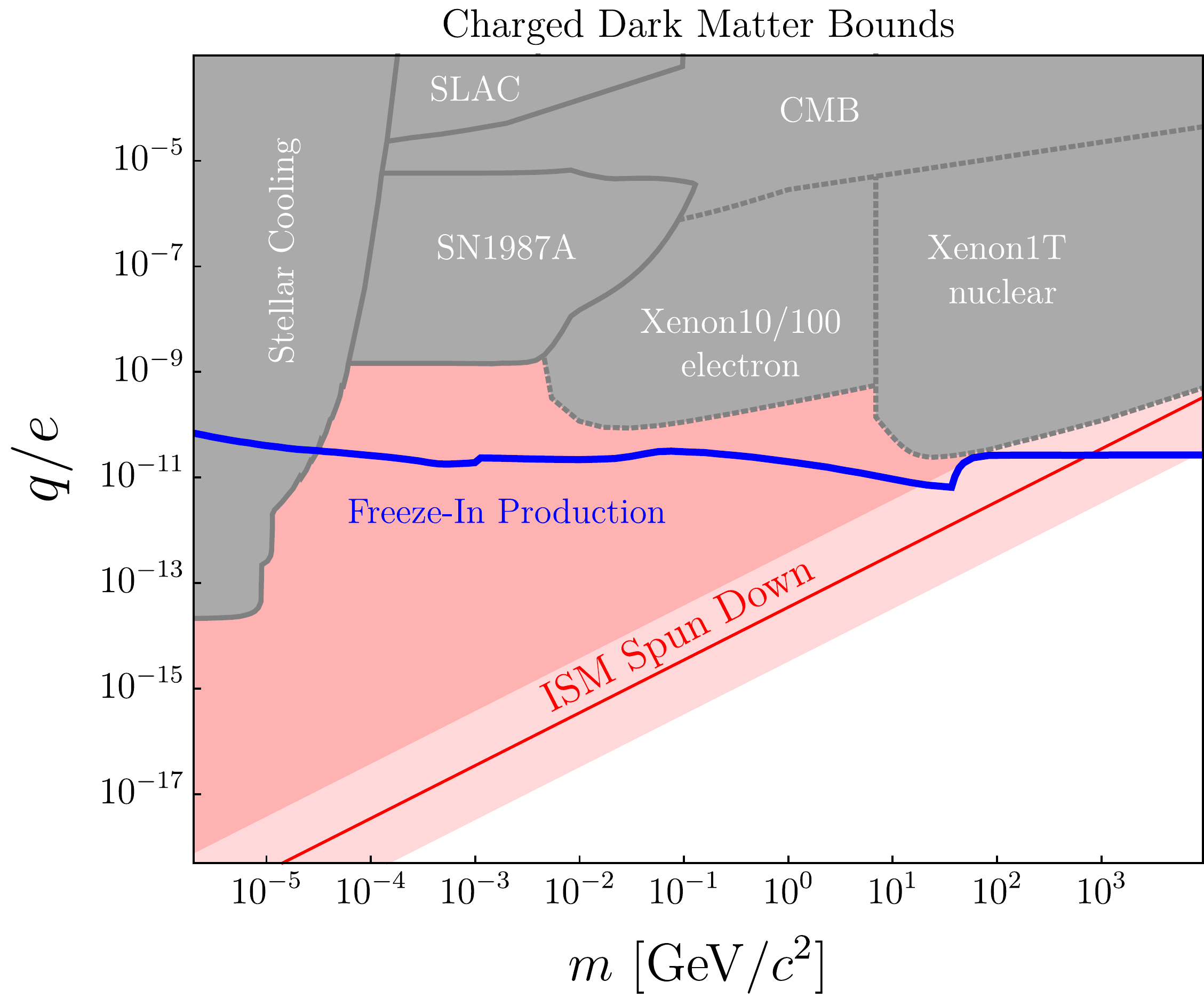}~~~~
\caption{ Constraints on ISM spin down assuming that qDM constitutes all of the halo DM $f_{\rm dm} = 1$. Here the ISM exclusion region is 
represented by the pink shaded region, which shows the 
the bound from Eq.~(\ref{local-spin-down-bound}) with the {\it ad hoc} uncertainties discussed in the text. The solid red curve is the 
central value based on local Galactic parameters and the pink band surrounding this line is an order-of-magnitude uncertainty band; we 
expect that the darker red shaded region is robustly excluded, but more detailed simulations are required to properly model the setup considered 
in this paper. The blue curve represents the parameter space for ``freeze-in" production of qDM in which their abundance is negligible at very early times (e.g. after inflation) and the present-day abundance arises entirely from annihilation of Standard Model particles in the universe after inflation and before matter-radiation equality \cite{Chu:2011be,Dvorkin:2019zdi} (also see Appendix \ref{appendix-freezein}).  Also shown are constraints on the existence of milli-charged particles by the SLAC millicharge experiment \cite{Prinz:1998ua}, supernova 1987a \cite{Chang:2018rso}, stellar cooling \cite{Vogel:2013raa}.  Additional constraints from CMB decoupling  \cite{Kovetz:2018zan,McDermott:2010pa}, Xenon10/100 (electron recoil direct detection) \cite{Essig:2012yx,Essig:2017kqs}, and Xenon1T (nuclear recoil direct detection) \cite{Hambye:2018dpi} assume $f_{\rm qdm} = 1$, so these regions are represented by dotted boundaries; these regions
also contain direct detection bounds from the qDM  sub-population accelerated by supernova shock waves \cite{Dunsky:2018mqs}.  
}
\label{millichargeBounds}

\end{figure*}

The milli-charged ($q\ll e$) dark matter scenario has inspired considerable attention in recent years partly because  milli-charges might acquire their cosmic abundance only through feeble electromagnetic interactions, the so-called ``freeze-in" mechanism  \cite{Dodelson:1993je,Hall:2009bx,Chu:2011be,Dvorkin:2019zdi}. In this scenario, the post inflationary universe initially contains essentially no milli-charged particles because of their weak coupling to the inflaton or to the particles which initially constitute the post-inflationary radiation bath. However, these thermalized particles  annihilate to yield milli-charges with an abundance determined only by the latter's feeble electromagnetic coupling. The freeze-in number density depends mostly on $q/m$ and the mass density on $q$.  For a broad range of masses $q\sim 10^{-11}e$  yields the observed dark matter density for a broad mass range of masses -- also see Appendix \ref{appendix-freezein} for more details.  This $q$-$m$ parameter space that yields the observed abundance is a key target of a new experimental program to detect DM in terrestrial laboratories (see \cite{Battaglieri:2017aum} for a review).  Furthermore, even if milli-charges constitutes only a small fraction of the total DM, it can yield interesting cosmological implications.  Indeed, it has recently been shown that a subdominant milli-charge population at $z\sim 17$ can significantly cool baryons relative to the $\Lambda$CDM prediction and thereby explain the $3\sigma$  anomaly in 21 cm absorption reported by the EDGES collaboration~\cite{nature,nature2,Fialkov:2018xre} (see also Refs.~\cite{Tashiro:2014tsa,Munoz:2015bca}).



However milli-charged qDM has faced stringent  limits from a variety of terrestrial experiments and astrophysical observations. Independently of their present-day cosmic abundance, the parameter space of milli-charged particles is constrained by various accelerator searches \cite{Prinz:1998ua}, observed stellar cooling rates \cite{Vogel:2013raa}, and supernova 1987A \cite{newsam}.  If their cosmic abundance is appreciable compared to the total DM density, there are additional constraints from qDM-baryon scattering during the CMB era \cite{Kovetz:2018zan},  dark matter direct detection searches \cite{Essig:2012yx,Essig:2017kqs}, and dark matter self-interaction bounds \cite{Agrawal:2016quu}. 
 Nonetheless, much of the interesting qDM parameter space has remained unexplored, including charge-to-mass ratios that can accommodate freeze-in production for masses above a few keV.  The ISM spin-down bound presented in this paper does now probe part of the ``freeze-in" parameter space, when $m c^2\lesssim 1\,{\rm TeV}$ as illustrated in Fig.~\ref{millichargeBounds} which compares the spin-down bound with other limits on milli-charged particles.  Note that there may also be limits on qDM from the observed shapes of galaxy cluster halo profiles \cite{Kadota:2016tqq}, but like the bound derived in this paper,  detailed numerical simulations are necessary to fully understand the parameter space constrained by this argument -- see Sec. \ref{best} for a discussion.

For stable charged particles with mass greater than $\sim1\,{\rm TeV}$ particles near the minimum $q/m$ allowed by the spin-down bound would, in the simplest scenarios, predict a qDM mass density in excess of the observed dark matter density.   One might then argue that the constraint on qDM from the dark matter density provides a more stringent bound than spin-down.  
This argument does not apply for sufficiently large $m$ since non-thermal qDM production is Boltzmann suppressed by the maximum temperature reached after inflation.  There are a variety of non-electromagnetic mechanisms of producing particles with mass $\sim10^{13}\,{\rm GeV}$ just after the end of inflation with the correct abundance for dark matter \cite{Chung1998}.  Such particles are sometimes called supermassive dark matter or WIMPzillas and may or may not carry electric charge. Curiously the spin-down bound on $m$ for $q=e$ is $\sim10^{13}\,{\rm GeV}$ and spin-down therefore constrains WIMPzilla parameter space.  Note that there is a mass gap between $1\,{\rm TeV}$ and WIMPzillas where one would not expect stable qDM particles to exist because the predicted (freeze-in density) is too large under standard cosmological assumptions.

ISM spin-down provides a method of constraining both milli-charged and WIMPzilla dark matter.  If evidence for ISM spin-down is discovered this would also provide evidence for the existence of one of these two.  However the ISM spin-down phenomenon by itself does not distinguish between these two extremely different hypotheses.

Finally, although our treatment has assumed that the qDM population is charged under electromagnetism, the argument presented here may have implications for other related scenarios. For example, if electrically neutral DM couples to Standard Model (SM) matter through a so-called dark photon, both SM and dark photons can undergo kinetic mixing \cite{Okun:1982xi,Holdom:1985ag}.  If both photons are massless diagonalizing the kinetic terms give DM a charge under SM electromagnetism and SM matter a charge under dark electromagnetism.   However if the dark photon is massive ($m_{\gamma'}>0$) then in vacuum kinetic mixing results in SM matter coupling to the dark photon but no DM coupling to massless SM photons.  Hence the DM does not acquire an electric charge in this scenario.  However this no longer holds in a plasma which gives the SM photon an effective mass, 
\be 
 \label{plasma-mass}
m_\gamma^{\rm pl}\,c^2\equiv\hbar\,\omega_{\rm p}=\hbar\,\left( \frac{4\pi\,n_{\rm e}\,e^2}{m_{\rm e}} \right)^{1/2 } 
= 7.4 \times 10^{-12}\,{\rm eV}\,\left( \frac{n_{\rm e}}{0.04 \, \rm cm^{-3}} \right)^{1/2}\ ,
\ee
 set by the plasma frequency, $\omega_{\rm p}$, where the $n_{\rm e}$ is the free electron density and $0.04$ cm$^{-3}$ is its typical value the solar neighborhood (see \cite{Dubovsky:2015cca} for a  discussion of kinetic mixing in a plasma).  If $m_{\gamma'}\ll m_\gamma^{\rm pl}$ then the DM acquires an effective millicharge under SM electromagnetism in spite of a finite dark photon mass.  A detailed analysis of such a scenario is beyond the scope of this paper but we expect that our conclusions about qDM should also apply to electrically neutral DM which couples to SM particles through a sufficiently light dark-photon.

\section{Conclusion}
\label{conclusion}

In this paper we have introduced a new bound on charged dark matter (qDM) derived from interactions with spiral galaxy interstellar medium (ISM) via their embedded ordered magnetic fields.  We have derived general formulae for the momentum transfer between a rotationally supported ISM disk and the more slowly rotating qDM halo passing through the disk.  Applying this result to the local ISM, whose $\sim$ kpc scale $\sim \mu$G magnetic fields are believed to have persisted for $\gtrsim $ 5 Gyr, we estimate that, in order to avoid {\it spinning-down} the ISM disk, the qDM charge-to-mass ratio must satisfy the prohibitive constraint
\be
\label{final}
\frac{q}{m\,c^2} \lesssim 3.5 \times 10^{-13\pm1} \left(\frac{e}{\rm GeV}\right)~,~~
\ee
if the qDM halo density fraction is $f_{\rm qdm} = \rho_{\rm qdm}/\rho_{\rm dm}  = 1$. However, for smaller halo fractions, $f_{\rm qdm} \lesssim 0.1$, there is insufficient qDM inertial mass passing through the disk to cause appreciable spin-down on Gyr timescales; instead the qDM population will be {\it spun-up} by the ISM disk. In this latter scenario, our observation does not imply a constraint, but does predict that the local qDM velocity distribution will differ from that of the dominant halo DM species. Note that in both the  ISM spin-down and qDM spin-up regimes, the qDM particles  participating in these momentum transfers receive net angular momentum; the only difference is whether the dominant qualitative changes take place in the ISM disk or in the qDM halo. 

We emphasize that our approach has made several key assumptions regarding the local Galaxy.  Namely, we have assumed that the presence of a qDM population at early times does not spoil our conventional understanding of galaxy formation. This assumption is required to justify our initial condition in which the qDM halo begins to extract angular momentum from the rotating ISM disk, which was not significantly different from its present form a few Gyr ago.  Our quantitative bound was based solely on the properties of the Solar neighborhood and required some choices in interpretation of observations of the local magnetic field which may have over- or underestimated the effect.  Setting the bound on $q/m$ from the local part of our Galaxy is conservative in the sense that there are other parts of our Galaxy and other galaxies which would likely yield even tighter bounds.  

While we have shown that qDM interactions with disordered small scale ISM magnetic fields is sub-dominant in the Born approximation, a more complete approach is called for which includes the larger field strength of these small scale fields as well as magnetic fields in galactic haloes which we have totally neglected.   To improve our argument, it is necessary to improve galactic magnetic field maps (both disordered and ordered components) and to perform numerical simulations to better understand the effect of qDM on the ISM formation history.  These important steps forward are beyond the scope of the present work, but are extremely well motivated in light of the strong qDM bound that our simple analysis implies. 

Indeed, the region of qDM parameter space covered by this bound is particularly compelling  from a theoretical perspective. At face value, the inequality in \Eq{final} covers parameter space predicted by cosmological {\it freeze-in} production of qDM for masses below a $\sim$ TeV$/c^2$.  In this scenario,   qDM is not initially produced during reheating after inflation, but is subsequently populated from annihilations of Standard Model, charged particles through their electromagnetic interactions. Unlike in the more familiar freeze-out scenario the DM production rate through these reactions is never sufficiently large for DM to thermalize with normal (Standard Model) matter. This freeze-in parameter region, $q/e\sim 10^{-11}$, is currently the focus of a broad experimental effort to detect millicharged particles in direct detection experiments (see Sec. 4 of  \cite{Battaglieri:2017aum} for a review).  Furthermore, even though our argument here has assumed that DM couples directly to the photon, our conclusions should also apply electrically neutral DM that couples to Standard Model particles through an ultralight, kinetically-mixed dark photon whose mass is much smaller than the typical ISM plasma mass acquired by the Standard Model photon.

We have focused only on placing limits on qDM parameters from spin-down on the assumption that such a dramatic difference between the ISM and stellar disks would not have escaped detection. However, we have not proposed an observational test of this phenomenon.  A more rigorous and precise bound could be obtained if such a test were to be devised.  Indeed, it is conceivable that evidence for spin-down might already be available in existing data or from future measurements of galactic properties; not necessarily in our local environment. In the event of a verified spin-down observation, it may be possible to measure the qDM $\sqrt{f_{\rm qdm}}\,m/q$ ratio or aspects of the DM velocity distribution.  This would happen near the boundary of the exclusion region of Fig.~\ref{qDMlimits} which include WIMPzillas and qDM which makes only a fraction $\sim0.2$ of the dark matter. One would want to corroborate spin-down evidence for qDM with other measurements and spin down would give target parameters for such measurements.  For example direct detection with Xenon 1T nuclear in the TeV mass range (see Fig.~\ref{millichargeBounds}) might overlap the boundary of our exclusion region.  Fully charged ($q=e$) ``WIMPzillas" near our exclusion boundary would pass though a tabletop sized detector a few times a year.   With proper shielding and background reduction this might also be detectable.

%

We emphasize that whether one is setting bounds or making discoveries one should first develop an observational test for spin-down which would allow one to quantify both statistical and systematic errors.  This is an important future step in validating the conclusions of this paper and putting more certain constraints on qDM.

\begin{acknowledgments}
{\it Acknowledgments} We thank Asher Berlin, Nikita Blinov, Nick Gnedin, Sam McDermott, and Diego Redigolo, and Christine Simpson for helpful conversations. Fermilab is operated by Fermi Research Alliance, LLC, under Contract No. DE-AC02-07CH11359 with the US Department of Energy. 
\end{acknowledgments}

\bibliography{MillichargeDiskDraftASGK5}

\begin{appendix}

\section{Generating Small Charges}
\label{appendix-small-charges}
Given the apparently quantized charges in the Standard Model (SM) of particles and interactions,  naively it seems ad-hoc to invoke  a much smaller charge $q \ll e$ for the dark matter.   However, such a scenario does, indeed arise without spoiling charge quantization if DM starts off uncharged under SM electromagnetism but couples to its own massless``dark photon" $A^\prime$.  In natural units $(c = \hbar = 1)$, the most general renormalizable and gauge-invariant Lagrangian contains
\be
{\cal L}_{\rm em + dm} = - \frac{1}{4} \left( F_{\mu\nu}^2 +  {F^\prime_{\mu\nu}}^{\!\! \!\! 2}   \right) +  \frac{ \kappa}{2} F^\prime_{\mu \nu} F^{\mu \nu} + e A_\mu J_{\rm em}^\mu + e^\prime A^\prime_\mu J_{\rm dm}^{\mu},
\ee
where $J_{\rm em}$ and $J_{\rm em}$  are, respectively,  the SM and dark electromagnetic current, $e^\prime$ is the DM coupling to $A^\prime$, and $\kappa$ is a dimensionless ``kinetic mixing" parameter.  Since $F$ and $F^\prime$ are separately gauge invariant, there is no symmetry forbidding kinetic mixing.  One can restore canonical normalization by shifting the dark photon $A^\prime \to A^\prime + \kappa A$ to yield 
\be
{\cal L}_{\rm em} \to - \frac{1}{4}  F_{\mu\nu}^2 +   A_\mu \left( e J_{\rm em}^\mu  +  q J_{\rm dm}^\mu \right) ,~~~ q \equiv \kappa e^\prime ~.
\ee
Thus, in this simple setup, DM acquires an electric charge $q$ under SM electromagnetism. A priori, there is  no preferred value of $\kappa$, but this parameter is expected to be very small if it arises radiatively from Feynman diagrams containing loops of heavy particles  that couple to both $A$ and $A^\prime$ \cite{Okun:1982xi,Holdom:1985ag}. However If the SM is described by a non-abelian unified gauge group at very high energies, kinetic mixing with a new abelian field cannot arise at tree level because non-abelian field strength tensors are not gauge invariant by themselves.  However, such a mixing can still arise when this non-abelian group is broken down to at least one abelian subgroup, whose field-strength can now mix with the dark abelian field.  Since there is no quantization requirement on $\kappa$, this mechanism can realize any value for the DM's charge under SM electromagnetism without spoiling any potential charge quantization requirement for either dark or SM sectors. 

If this model is extended to allow for a nonzero dark photon mass, then it is no longer true that the DM acquires charge under SM electromagnetism; such a coupling can be diagonalized away. However, as observed in Ref. \cite{Dubovsky:2015cca} and discussed in Sec. \ref{models}, in the ISM, the SM photon acquires a plasma mass of order $\sim 10^{-12}$ eV, so when in-medium effects are included, 
the dark matter will acquire a coupling to the SM photon if the dark photon is lighter than this value.

\section{Cosmological Freeze In Production}
\label{appendix-freezein}
In this appendix we summarize the freeze in production mechanism under standard cosmological assumptions following the treatment in \cite{Chu:2011be}. After inflation ends and the universe is  radiation dominated, we assume that the initial DM density is negligible and that its late-time population arises only from its interaction with the hot radiation bath of Standard Model (SM) particles prior to matter-radiation-equality.  In this appendix we use natural units: $\hbar = c = k_{\rm B}= 1$, there is only SM electromagnetism and $\alpha$ is the SM fine structure constant.

In a Friedman-Lemaitre-Robertson-Walker metric, the Boltzmann equation for qDM production and annihilation via $f^+f^- \leftrightarrow$ qDM qDM  reactions is 
\be
\frac{dn}{dt} + 3 H n = - \langle \sigma v \rangle \left[ n^2 - n_{\rm eq}^2   \right]~~~,~~~
H = \frac{\dot a}{a},
\ee
where $a$ is the cosmic scale factor, $\langle \sigma v\rangle$ is the thermally averaged annihilation cross section times velocity, $n$ is the number density of qDM particles in the universe, $n_{\rm eq}$ is their thermal equilibrium density, $H$ is the Hubble expansion rate, and $f$ is any SM charges particle. During radiation domination and for $T \gg m$ the latter two quantities can be written in terms of the temperature 
 \be
 H(T) = 1.66 \sqrt{g_\star(T)}\frac{T^2}{M_{\rm Pl}} ~~~~,~~~~n_{\rm eq}(T) = \frac{3 \zeta(3)}{\pi^2} T^3~~~,
 \ee
where $g_\star$ is the effective number of relativistic degrees of freedom and $M_{\rm Pl} = 1.22 \times 10^{19}$ GeV is the Planck mass.  We have assumed that the qDM particle is a Dirac fermion. .
 
If DM is not initially produced when inflation ends and the universe first becomes radiation dominated, then the initial qDM density satisfies $n(t=0) = 0$.  If we also demand that  $q \ll e$ so that the DM annihilation rate is much slower than the Hubble rate throughout its production then we can neglect the $n^2$ term  in the Boltzmann equation which now becomes integrable
\be
\frac{dn}{dt} + 3 H n = s \frac{dY}{dt} =  \langle \sigma v  \rangle (n_{\rm eq} )^2  ~~~,~~~ Y = \frac{n}{s} ~~,
\ee
where $Y$ is the comoving qDM yield, $s = 2 \pi^2 g_{\star, s} T^3/45$ is the entropy density in radiation domination and $g_{\star, s}$ is the number of relativistic degrees of freedom in entropy.  Assuming Maxwell-Boltzmann statistics the thermally averaged cross section can be written as a simple function of the temperature \cite{Gondolo:1990dk}
\be
\langle \sigma v \rangle =
 \frac{1}{8\,m^4\,T\,\left(K_2\left(\frac{m}{T} \right)\right)^2} \int_{4m^2}^\infty  ds\,\sigma(s)\sqrt{s}\,(s - 4m^2)\,K_1\left(\frac{\sqrt{s}}{T} \right)
 ~~,~~ 
\sigma(s) = \frac{4\,q^2\,q_f^2\,\alpha^2}{3s},
\ee
where $s$ is the mandelstam variable, $K_i$ is the order $i$ modified Bessel function of the 2nd kind, $\sigma(s)$ is the $ff \to $ qDM qDM cross section in the ultra relativistic limit, $q$ is the qDM charge, and $q_f$ is the Standard Model charge of the Standard Model particle $f$.  Since $T \propto a^{-1}$ we can swap time for temperature and integrate to obtain the asymptotic qDM yield at late times
\be
\Omega_{\rm qdm} = \frac{\rho_{\rm qdm}}{\rho_{\rm crit}} = \frac{m_\chi s_0}{ \rho_{\rm cr}}    \int_\infty^{m_\chi} \frac{dT}{T} \frac{   n_{\rm eq} ^2 }{H s}    \langle \sigma v \rangle~,~~
\ee
where $s_0$ is the present day CMB entropy, $\rho_{\rm crit}$ is the present day critical cosmological density and we have used comoving entropy conservation to obtain the final density $\rho_\chi = m_\chi n_\chi = m_\chi Y_f s_0$. For most of the mass range shown in Fig. \ref{millichargeBounds}, the observed DM abundance $\Omega_{\rm qdm} \sim 0.2$ is achieved for fractional charges of order  $q/e \sim 10^{-11}$. Note that the calculation outlined in this appendix does not take into account plasmon mixing effects that modify the production rate for millicharged particles below the electron mass \cite{Dvorkin:2019zdi}; however these effects are included in the blue freeze-in curve in Fig.~\ref{millichargeBounds}. 
 
\end{appendix}

\end{document}